\documentclass[peerreview]{IEEEtran}

\usepackage[square,numbers,sort&compress]{natbib}
\usepackage{hyperref}
\hypersetup{
    colorlinks,
    linkcolor={blue!80!black},% 
    citecolor={green!50!black},
    urlcolor={blue!80!black}
}

\usepackage{amsthm}
\usepackage{amsmath}
\usepackage{mathrsfs}
\usepackage{amssymb} 
\usepackage{stackrel}
\usepackage{mathtools}
\usepackage{nicefrac}

\usepackage{physics}

\usepackage{verbatim}
\usepackage{enumerate}

\usepackage[T1]{fontenc}
\usepackage{bbm} 
\usepackage{bbding}
\usepackage{keystroke}
\usepackage{pifont}
\usepackage{relsize}

\usepackage{graphicx}
\usepackage{xcolor}
\usepackage{tikz}
\usetikzlibrary{calc}
\usepackage{tabularx}
\usetikzlibrary{arrows,shapes}

\renewcommand{\trace}{\mathrm{Tr}}

\newcommand{\identity}{\mathbbm{1}}

\newcommand{\markovC}[1]{% 
\begin{tikzpicture}[#1]% 
\draw (0,0.3ex) -- (1ex,0.3ex);% 
\draw (0.5ex,0.3ex) circle (0.2ex);
\draw[white] (0.2ex,0) -- (0.5ex,0);% 
\end{tikzpicture}% 
}
\newcommand{\Cbar}{\markovC{scale=2}}

\theoremstyle{remark}	\newtheorem{theorem}{Theorem}
\theoremstyle{remark}	\newtheorem{lemma}[theorem]{Lemma}
\theoremstyle{remark}	
\theoremstyle{remark}	
\theoremstyle{remark} \newtheorem{definition}{Definition}
\theoremstyle{remark} \newtheorem{remark}{Remark}
\theoremstyle{remark} \newtheorem{example}{Example}

\title{Classical Feedback in a Quantum Network}

\author{ 
    \IEEEauthorblockN{Elina Levi and Uzi Pereg} \\
    \IEEEauthorblockA{\normalsize
        ECE Department \& Helen Diller Quantum Center, Technion \\
        {\tt elina.levi@campus.technion.ac.il, uzipereg@technion.ac.il}
    }
}

\date{\today}

\begin{document}
\maketitle

\begin{abstract}
Communication over a quantum multiple access channel (MAC) is considered with classical feedback. Since the no-cloning prohibits universal copying of arbitrary quantum states, classical feedback is generated through measurement. An achievable rate region is derived using partial information decoding at each transmitter. Our region generalizes both the classical Cover–Leung region and the generalized feedback region. As an example, we show that the qubit SWAP channel can benefit from feedback.
\end{abstract}

\section{Introduction}

The development of quantum communication has attracted increasing attention as quantum technologies advance toward real-world deployment \cite{Liu:25p}. 
These advances will enable a range of applications such as
unconditional security \cite{Cao:22p}, distributed quantum computing \cite{Caleffi:24p}, and quantum sensing \cite{Aslam:23p}.
By contrast, classical communication systems are already highly efficient and reliable, with modern standards such as 5G incorporating capacity-achieving codes  \cite{Vcarapic:20p}. 
A promising direction for near-term implementation is the integration of quantum capabilities into existing classical infrastructures, giving rise to hybrid classical-quantum networks
\cite{LukensPetersQi:25a}.
Within this framework, it is also natural to design systems that leverage classical resources to support and enhance quantum networks. In particular, the incorporation of classical feedback into quantum multi-user networks is of special interest.

A key question is how cooperative resources can improve performance in multi-user networks. 
Recent work considered cooperative quantum resources in relay \cite{Pereg:25p, IlinPereg:25a}, interference \cite{Hawellek:25p}, and broadcast channels \cite{PeregDeppeBoche:21p, FawziFerme:24p, YaoJafar:25p}.
One of the most fundamental network models in multi-user information theory is the multiple access channel (MAC) \cite{Liu24:p}. 
In the classical setting, recent studies have examined the advantages of entanglement assistance between transmitters \cite{Leditzky:20p, PeregDeppeBoche:25p} 
 and non-signaling assistance \cite{FawziFerme:23p}.
In the quantum setting, 
the capacity region of the quantum MAC without additional resources has been characterized in a regularized form for classical information transmission \cite{Winter:01p}. 
Recently considered cooperative resources include entanglement assistance between transmitter and receiver \cite{HsiehDevetakWinter:08p}, between transmitters \cite{LiuChenXi:25p},
conferencing links between transmitters \cite{BocheNotzel:14p}, 
and cribbing side information at one of the transmitters \cite{PeregDeppeBoche:22p}. 
In interactive communication systems, feedback is naturally available and can serve as a resource for cooperation.

The role of feedback has been extensively studied in classical information theory \cite{Xu:24p}.
In the classical model with feedback, the transmission $X_i$ at time $i$ can be a function of the message and of the past channel outputs $Y_1,\ldots, Y_{i-1}$.
Feedback can thus be viewed as the receiver providing a copy of the received output to the transmitter, through a back channel.
For classical single-user memoryless channels, feedback does not increase capacity,
though it can enhance the zero-error capacity \cite{Shannon:56p}, \cite[Sec. 3.9]{Kramer:08b}.
For channels with memory, however, feedback is known to increase capacity \cite{CoverPombra:89p}.
In the quantum case, Bowen et al. \cite{Bowen:05p} showed that classical feedback does not improve the capacity of a single-user memoryless quantum channel. 
The no-cloning theorem, a fundamental result of quantum mechanics, prohibits universal copying of an arbitrary quantum state. Thus, sending a copy of the quantum receiver's output state is impossible in general. Instead, classical feedback provides a noiseless classical back channel from receiver to transmitter.

Remarkably, feedback can increase communication rates for a classical MAC, even in the memoryless model. This effect was first demonstrated by Gaarder and Wolf via the binary adder MAC \cite{GaarderWolf:75p}, and later extended to an achievable rate region by Cover and Leung \cite{CoverLeung:81p}, with further improvements by Bross and Lapidoth \cite{BrossLapidoth:05p}, and by Venkataramanan and Pradhan \cite{VenkataramananPradhan:11p}. The multi-letter characterization involves the directed information \cite{Kramer:98t} (see also \cite{PermuterWeissmanChen:09p}).
Intuitively, the Cover-Leung inner bound is tight when one transmitter can perfectly recover the other’s message from the channel output \cite{Willems:82p}.
The Gaussian MAC provides an example where the Cover-Leung inner bound is \emph{not} tight  \cite{Ozarow:84p}. 
Variants of their model have also been studied, including generalized feedback \cite{Carleial:82p}, imperfect feedback \cite{LapidothWigger:10p},
and rate-limited feedback \cite{ShavivSteinberg:13p}.
Feedback has proven beneficial in a variety of scenarios, including the Gaussian broadcast channel \cite{OzarowLeung:84p, Gastpar:14p}, and secure communication \cite{LaiElGamalPoor:08p, Ardestanizadeh:09p, VenkataramananPradhan:13p, BassiPiantanidaShamai:19p}. Closely related are models in which causal side information is available at the transmitter (CSIT), which have been investigated in general settings \cite{CaireShamai:99p, RosenzweigSteinbergShamai:05p}, for broadcast channels \cite{SteinbergWeissman:12p}, and in combination with feedback \cite{MerhavWeissman:06p}.
In practical applications, feedback also serves as a valuable coding tool, reducing both encoding and decoding complexity while significantly lowering block error rates. A common approach to constructing feedback codes is to concatenate an inner and an outer code. Examples include concatenations of LDPC codes with belief-propagation and closed-loop iterative doping algorithms \cite{CaireShamaiVerdu:05p}, and linear codes for AWGN channels with noisy feedback \cite{MishraVasalKim:23p}. New designs include variable-length sparse feedback codes \cite{YavasKostinaEffros:24p} and binary error-correcting codes with feedback \cite{GuptaGuruswamiYunZhang:25p}.
Insights from the classical setting motivate studying how feedback effects can enhance communication in quantum MACs.

In this work, we study the advantage from classical feedback in communication over a quantum MAC, as illustrated in Fig.~\ref{figure:quantum_MAC_classical_feedback}. In particular, we show that for the qubit SWAP channel \cite{KlimovitchWinter:05a}, feedback increases achievable rates. We further establish an achievable rate region for the general memoryless quantum MAC with classical feedback. Specifically, the  receiver generates feedback through a quantum measurement on the channel output, collapsing it to a classical outcome that is sent back to the transmitters. The combination of the post-measurement system and measurement outcome constitutes the information available for decoding the messages.
These results demonstrate that classical feedback can enhance quantum multi-user communication and provide a foundation for further study of feedback-assisted quantum networks.

To prove achievability, we combine quantum information-theoretic tools with the classical feedback approach to construct a code for the quantum MAC with feedback.
Specifically, we adapt the classical block Markov coding scheme using a three-layered superposition code with backward decoding to the quantum setting. 
We use $T$ transmission blocks, each consists of $n$ channel uses, to send a sequence of messages. In our scheme, the transmitters employ a partial decoding strategy in which each transmitter decodes part of the other’s message from the previous block using its feedback, and leverages this information to encode cooperatively.
We consider a full decoding strategy, where each transmitter decodes the other's entire message, and demonstrate that partial decoding can be strictly better than full decoding.
We use the quantum multiparty packing lemma due to Ding et al. \cite{Ding:20p}, which generalizes the classical counterpart to a multiplex Bayesian network.
In our achievability scheme, the receiver applies a decoding measurement that recovers both messages from the same quantum state, within each block of our layered scheme.

The structure of the paper is as follows.
In Section~\ref{section:definitions_channel_model} we introduce the necessary preliminaries and the model of the quantum MAC with classical feedback.
In Section~\ref{section:coding}, we develop the coding framework for this setting, where the feedback is generated through measurement.
Section~\ref{section:main} establishes our main result, an achievable rate region, and the example of the qubit SWAP channel.
We conclude the paper in Section~\ref{section:summary}. The quantum multiparty packing lemma is described at Appendix~\ref{appendix:quantum_multiparty_packing_lemma}, the proof outline for Theorem~\ref{theorem:quantum_cover_leung} in Appendix~\ref{appendix:quantum_cover_leung_proof_outline}, 
and the proof for Theorem~\ref{theorem:quantum_inner_bound} in Appendix~\ref{appendix:quantum_inner_bound_proof}.

\section{Definitions and Channel Model}
\label{section:definitions_channel_model}
\subsection{Notation}
We use the following notation conventions; $\mathcal{X}, \mathcal{Y}$ are used for finite sets, $X,Y$ for random variables, and $x,y$ are used for their realizations. 
Vectors of length $n$ are denoted in bold, such as $\mathbf{x}$.
Given a random variable $X \sim p_X(x)$, the $\delta$-typical set is defined as
\begin{align}
    \mathcal{T}^{(n)}_\delta(X)&= \left\{ \mathbf{x} \in \mathcal{X}^n : |\pi(x|\mathbf{x})-p_X(x)|\leq \delta \cdot p_X(x) \text{ for all } x\in \mathcal{X} \right\}
    \intertext{where}
    &\pi(x|\mathbf{x}) = \frac{|\{i:x_i=x\}|}{n} \,.
\end{align}

\subsection{Quantum States, Measurements, and Instruments}
A quantum state is represented by a density matrix, denoted by $\rho\in\mathfrak{D}(\mathcal{H})$, where $\mathfrak{D}(\mathcal{H})$ is the set of all density operators on the Hilbert space $\mathcal{H}$.
A generalized measurement is specified by a set of operators, $\{D_z\}$, such that $\sum_z D_z^\dagger D_z=\identity$. According to the Born rule, the probability of the measurement outcome $z$ is $\Pr(z)=\trace\{D^\dagger_zD_z\rho\}$ and the post-measurement state is $\rho'=\frac{D_z\rho D_z^\dagger}{P_Z(z)}$.
The distribution of measurement outcomes can also be described in terms of 
a positive operator-valued measure (POVM), i.e.,  a set $\{\Delta_z\}_z$ of operators such that
$\sum_z\Delta_z=\identity$, where $\identity$ is the identity operator.

A quantum channel is a linear, completely-positive trace-preserving (CPTP) map, denoted by $\mathcal{N}_{A\to B}$.
A quantum instrument $\Lambda$ is a quantum channel constructed from a collection $\{\Lambda_z\}$ of completely positive, trace non-increasing maps, such that $\Lambda(\rho)=\sum_z \Lambda_z(\rho) \otimes \ketbra{z}_Z$, where $\{\ket{z}\}$ is an orthonormal basis for a Hilbert space $\mathcal{H}_Z$.

A quantum instrument can be used to describe a measurement on a quantum system $B$, where the output includes both the classical measurement outcome $\ketbra{z}$ and the post-measurement quantum system $\Lambda_z(\rho_B)=\rho_{\bar{B}}^{(z)}$. It can be constructed from a generalized measurement $\{D_z\}$ by defining the completely positive, trace-non-increasing maps $\Lambda_z$ as $\Lambda_z(\rho) = D_z \rho D_z^\dagger$ and
\begin{align}
    \Lambda_{B\to \bar{B}Z}(\rho_B)=\sum_z D_z\rho_B D_z^\dagger \otimes \ketbra{z}_Z
    =\sum_z \Pr(z)\rho_{\bar{B}}^{(z)} \otimes \ketbra{z}_Z \,.
\end{align}

\subsection{Quantum Information Measures}
The quantum entropy of $\rho\in\mathfrak{D}(\mathcal{H})$ is defined as 
\begin{align}
    H(\rho)=-\trace(\rho\log\rho) \,.
\end{align}
For a bipartite state $\rho_{AB} \in \mathfrak{D}(\mathcal{H}_A \otimes \mathcal{H}_B)$, the quantum mutual information is
\begin{align}
    I(A;B)_\rho = H(\rho_A) + H(\rho_B) - H(\rho_{AB})\,.
\end{align}
The conditional quantum entropy and mutual information are defined by
$H(A|B)_\rho=H(\rho_{AB})-H(\rho_B)$ and $ I(A;B|C)_\rho~=~H(A|C)_\rho + H(B|C)_\rho - H(A,B|C)_\rho$, respectively. \\

The Holevo information of a quantum channel $\mathcal{N}_{A\to B}$ is defined as \cite[Sec. 13.3]{Wilde:11b}
\begin{align}
    \chi(\mathcal{N}) = \max_{\{p_X(x),\rho_x,\Delta\}} I(X;B)_\rho
\end{align}
where the maximization is over all input ensembles
$\{p_X(x),\rho_x\}_{x\in\mathcal{X}}$ in
$ \mathfrak{D}(\mathcal{H}_A)$,  hence the classical-quantum state $\rho_{XB}$ is given by
\begin{align*}
\rho_{XB}=\sum_{x\in\mathcal{X}} p_X(x) \ketbra{x}_X\otimes \mathcal{N}_{A\to B}(\rho_x) \,.
\end{align*}
For several classes of channels, the capacity is given by the Holevo information. Specifically, this holds for classical-quantum channels, among other examples 
\cite[Sec. 20.4]{Wilde:11b}.

The accessible information of a quantum channel $\mathcal{N}_{A\to B}$ is the coding rate that corresponds to the naive approach of decoding by a product of measurements 
$\{\Delta_z^{\otimes n}\}$ on the channel output \cite[Sec. 20.1]{Wilde:11b}. 
The accessible information
is defined as
\begin{align}
    I_{\text{acc}}(\mathcal{N}) = \max_{\{p_X(x),\rho_x,\Delta\}} I(X;Z)
\end{align}
where the maximization is over all input distributions $p_X(x)$, collections of input density operators $\{\rho_x\}\subset \mathfrak{D}(\mathcal{H}_A)$, and POVMs $\Delta=\{\Delta_z\}$, hence the random pair $(X,Z)$ is distributed according to 
\begin{align*}
p_{XZ}(x,z)=p_X(x) \trace\left[ \Delta_z\cdot \mathcal{N}_{A\to B}(\rho_x) \right] \,.
\end{align*}

\subsection{Quantum Multiple-Access Channel}
We consider the communication task of sending messages via a fully quantum MAC $\mathcal{N}_{A_1A_2 \to B}$ with the assistance of a classical feedback link, see Fig.~\ref{figure:quantum_MAC_classical_feedback}, where $A_1$ and $A_2$ represent  Transmitter 1 (``Alice~1'') and Transmitter 2 (``Alice~2''), respectively, $B$ is the receiver (``Bob''), and $Z_1$ and $Z_2$ are the classical feedback messages to Alice~1 and Alice~2, respectively.
We note that in the classical model with feedback,  Bob provides a copy of his received output % 
through a back channel. This allows the transmitters to obtain information about the other's message, enabling cooperation.
In the quantum setting, the no-cloning theorem \cite[Sec. 3.5.4]{Wilde:11b} prohibits perfect copying.  
Instead, feedback will be defined through measurement, see Section~\ref{section:coding}.

\section{Coding With Classical Feedback}
\label{section:coding}
\begin{figure}[tb]
\begin{center}
\includegraphics[width=0.5\linewidth]{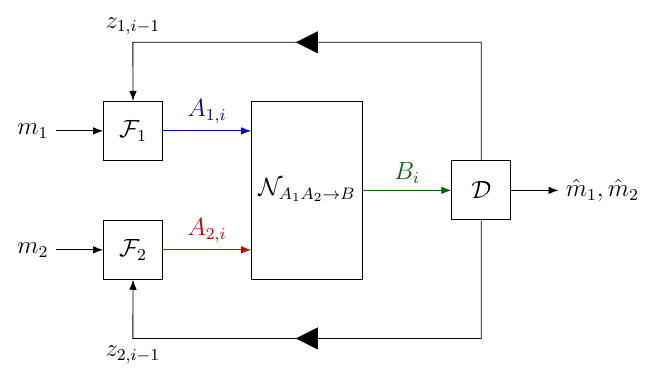}
\end{center}
\caption{Quantum MAC with classical feedback.}
\label{figure:quantum_MAC_classical_feedback}
\end{figure}
Consider a quantum MAC $\mathcal{N}_{A_1,A_2\to B}$. We define
a code for the transmission of messages via $\mathcal{N}_{A_1,A_2\to B}$ with feedback, i.e., noiseless classical links from Bob to Alice~1 and from Bob to Alice~2 (see Fig.~\ref{figure:quantum_MAC_classical_feedback}).
Through feedback, the transmitters can gain partial knowledge of each other's message, which becomes shared knowledge, and allows them to cooperate.

\begin{definition}[Code with classical feedback] A $(2^{nR_1},2^{nR_2},n)$  code for the quantum MAC $\mathcal{N}_{A_1,A_2\to B}$ with classical feedback consists of the following:
\begin{itemize}

\item Two message sets $\mathcal{M}_1=[1:2^{nR_1}]$ and $\mathcal{M}_2=[1:2^{nR_2}]$, for Alice~1 and Alice~2, respectively, where $2^{nR_k}$ is assumed to be integer.

\item A sequence of strictly casual encoding maps 
$\mathcal{F}_k = \left\{\mathcal{F}_{M_kZ_{k}^{i-1}\to A_k^i}^{(k,i)}:\mathcal{M}_k \times \mathcal{Z}_k^{i-1}\to  \mathfrak{D}(\mathcal{H}_{A_k}^{\otimes i})\right\}_{i \in [1:n]}$
for Alice $k$, where $k\in\{1,2\}$, such that each encoding map
is backward compatible.
That is, the joint input state:
\begin{align}
    \rho_{A_k^i}^{(m_k,z_k^{i-1})} &= 
    \mathcal{F}_{M_k Z_k^{i-1}\to A_k^i}^{(k,i)}\left(m_k,z_{k}^{i-1}\right) \\
    \intertext{must satisfy}
    \rho_{A_k^{i-1}}^{(m_k,z_k^{i-2})} &= \trace_{A_{k,i}}\left(\rho_{A_k^i}^{(m_k,z_k^{i-1})}\right).
    \intertext{Equivalently}
    \mathcal{F}_{M_k Z_k^{i-2}\to A_k^{i-1}}^{(k,i-1)} &= \trace_{A_{k,i}} \circ \mathcal{F}_{M_k Z_k^{i-1}\to A_k^i}^{(k,i)} \nonumber \,.
\end{align}
We note that this requirement resembles that of causal side information \cite{Pereg:21p}.

\item A sequence of feedback quantum instruments $\mathcal{D} = \left\{\mathcal{D}^{(i)}_{B_{i}\bar{B}^{i-1}\to\bar{B}^{i}Z_{1,i} Z_{2,i}}\right\}
$, where $\bar{B}^{i}$ 
is the post-measurement system at time $i$.

\item A decoding POVM $\Delta = \left\{\Delta_{m_1,m_2 |z_1^{n-1},z_2^{n-1}}\right\}$ on $\mathcal{H}_{\bar{B}}^{\otimes (n-1)}\otimes \mathcal{H}_B$, producing a measurement outcome $(m_1,m_2)\in \mathcal{M}_1\times \mathcal{M}_2$. 
\end{itemize}

We denote the code by $(\mathcal{F}_1,\mathcal{F}_2,\mathcal{D}, \Delta)$.
\end{definition}

The coding scheme works as follows; Alice $k$ selects a message $m_k\in \mathcal{M}_k$, where $k\in \{1,2\}$.
At time $i$, Alice $k$ encodes the message with the encoding map 
$\mathcal{F}^{(k,i)}$
, using the feedback output $z_k^{i-1}$ that is available at time $i$. Alice~1 and Alice~2 then send $A_{1,i}$ and $A_{2,i}$, respectively, through the quantum MAC $\mathcal{N}_{A_1A_2\to B}$. Bob performs a measurement using the feedback quantum instrument $\mathcal{D}^{(i)}
$ on the channel output $B_i$ and the post-measurement system $\bar{B}^{i-1}$ from the previous step, and sends the measurement outcome $z_{1,i}, z_{2,i}$ through a feedback link that introduce a single-unit delay, to Alice~1,~2. Therefore,  at time $i$,  Alice~1 and Alice~2 get $z_{1,i-1}$ and $z_{2,i-1}$, respectively.

Specifically, at time $i=1$, Alice $k$ encodes its message $m_k$ using 
$\mathcal{F}_{M_k\to A_{k,1}}^{(k,1)}$.
Bob receives the channel output $B_1$, and applies the quantum instrument
$\mathcal{D}^{(1)}_{B_1\to \bar{B}_1 Z_{1,1} Z_{1,2}}$, obtaining the classical outcomes $Z_{1,1},Z_{2,1}$ along with the post-measurement  system $\bar{B}_1$. Bob sends $Z_{1,1}$ through the first feedback link to Alice~1,  and $Z_{2,1}$ through the second feedback link to Alice~2.

At time $i=2$, Alice $k$ has the feedback outcome $Z_{k,1}$.
Given $Z_{k,1}=z_{k,1}$, she
uses the encoding map 
$\mathcal{F}_{M_k Z_{k,1}\to A_k^2}^{(k,2)}(m_k,z_{k,1})$,
and sends $A_{k,2}$
through the channel.

Bob receives the channel output $B_2$ and similarly as before, he applies the quantum instrument $\mathcal{D}^{(2)}_{B_2\bar{B}_1\to \bar{B}^2 Z_{1,2} Z_{2,2}}$, obtaining the classical outcomes $Z_{1,2},Z_{2,2}$ along with the new post-measurement system $\bar{B}^2$. Bob sends $Z_{1,2}$ and $Z_{2,2}$ through the feedback links to Alice~1 and Alice~2, respectively. This continues in the same manner.

After time $i=n$,
Bob receives the output $B_n$ in the state
\begin{align}
\rho_{B_n \bar{B}^{n-1}}^{(m_1,m_2,z_1^{n-1},z_2^{n-1})} = 
(\mathcal{N}_{A_1A_2\to B}\otimes\mathrm{id}_{\bar{B}^{n-1}})\left(\rho_{A_{1,n} A_{2,n} \bar{B}^{n-1}}^{(m_1,z_{1}^{n-1},m_2,z_{2}^{n-1})} 
\right)
\label{Equation:Output_n}
\end{align}
where $\rho_{A_{1,n} A_{2,n} \bar{B}^{n-1}}^{(m_1,z_{1}^{n-1},m_2,z_{2}^{n-1})}$ is the input given 
$Z_k^{n-1}=z_k^{n-1}$. % 

To decode, Bob performs the final measurement $\Delta_{m_1,m_2 |z_1^{n-1},z_2^{n-1}}$ on $B_n\bar{B}^{n-1}$ and obtains an estimate $(\hat{m}_1,\hat{m}_2)$
of the messages. 

The conditional probability of error given
$(m_1,m_2,z_1^{n-1},z_2^{n-1})$, is
\begin{align}
   P_e^{(n)}(\mathcal{F}_1,\mathcal{F}_2,\mathcal{D}, \Delta |m_1,m_2,z_1^{n-1},z_2^{n-1})=1-\trace\left(\Delta_{m_1,m_2|z_1^{n-1},z_2^{n-1}}\rho_{B_n\bar{B}^{n-1}}^{(m_1,m_2,z_1^{n-1},z_2^{n-1})}\right)
\end{align}
where $\rho_{B_n\bar{B}^{n-1}}^{(m_1,m_2,z_1^{n-1},z_2^{n-1})}$ is as in \eqref{Equation:Output_n}.

The average probability of error of the code $(\mathcal{F}_1,\mathcal{F}_2,\mathcal{D}, \Delta)$, under the assumption that the messages are uniformly distributed, is given by
\begin{align}
P_e^{(n)}(\mathcal{F}_1,\mathcal{F}_2,\mathcal{D}, \Delta)=\frac{1}{|\mathcal{M}_1||\mathcal{M}_2|}\sum_{m_1=1}^{|\mathcal{M}_1|}\sum_{m_2=1}^{|\mathcal{M}_2|}\sum_{z_1^{n-1},z_2^{n-1}}&\Pr(Z_1^{n-1}=z_1^{n-1},Z_2^{n-1}=z_2^{n-1}|m_1,m_2) \nonumber \\
&\cdot P_e^{(n)}(\mathcal{F}_1,\mathcal{F}_2,\mathcal{D}, \Delta |m_1,m_2,z_1^{n-1},z_2^{n-1})
\end{align}
with
\begin{align}
    \Pr(Z_{1,i}=z_{1,i},Z_{2,i}=z_{2,i}|z_1^{i-1},z_2^{i-1},m_1,m_2)
    = \trace\left\{D_{z_{1,i} z_{2,i}}^\dagger D_{z_{1,i} z_{2,i}} \cdot \rho_{B_i\bar{B}^{i-1}}^{(m_1,m_2,z_1^{i-1},z_2^{i-1})}\right\}
  \end{align}
 where $D_{z_{1,i} z_{2,i}}$ correspond to the feedback measurement.

\begin{remark}[Operational description]
In practice, encoding with feedback can be implemented as follows.
 Alice $k$ first prepares a joint auxiliary state $\Psi_{\bar{A}_{k,1}\bar{A}_{k,2}\cdots\bar{A}_{k,n}}$. Then, at each time instance, % 
she applies an encoding map of the form $\mathcal{E}_{\bar{A}_{1,i}\to A_{1,i}}^{(m,z_1^{i-1})}$. 
\end{remark}

\begin{remark}[Communication without feedback]
\label{remark:communication_without_feedback}
By choosing the quantum instrument of the decoder's feedback to be the identity map, the model reduces to the standard quantum MAC without feedback \cite{Winter:01p}.
\end{remark}

\begin{remark}[Perfect feedback and generalized feedback]
In a classical MAC with \emph{perfect} feedback, the channel output $Y$ is sent through a noiseless link back to both transmitters.
That is, the feedback messages $Z_1$ and $Z_2$ are identical to the classical output $Y$ \cite{CoverLeung:81p}.
As pointed out earlier, perfect feedback is impossible in the quantum setting, due to the no-cloning theorem. 
In a classical MAC with \emph{generalized} feedback, the channel model is defined in terms of a % 
fixed probability function $P_{Z_1Z_2Y|X_1X_2}$ % 
\cite{Carleial:82p}.
In our model, 
the receiver is free to choose an arbitrary quantum instrument, and thus dictates the feedback statistics. We will see the implications in Theorem~\ref{theorem:quantum_inner_bound}.
\end{remark}

We now define achievable rates and the capacity region with classical feedback.  
\begin{definition}[Achievable rate pair]
A rate pair $(R_1,R_2)$ is 
achievable for the quantum MAC with classical feedback, if for every $\varepsilon, \delta>0$ and sufficiently large $n$, there exists a $(2^{n(R_1-\delta)},2^{n(R_2-\delta)},n)$ code such that $P_e^{(n)}(\mathcal{F}_1,\mathcal{F}_2,\mathcal{D}, \Delta)\leq \varepsilon$.   
\end{definition}

\begin{definition}[Capacity region]
The capacity region of the quantum MAC with classical feedback, denoted by $\mathcal{C}_{\text{fb}}(\mathcal{N})$, is the closure of the set of all achievable rate pairs.    
\end{definition} 

\section{Main Results}
\label{section:main}
We now present our main results for the quantum MAC $\mathcal{N}_{A_1A_2\to B}$ with classical feedback.
In particular, we establish two achievable rate regions for this setting.

\subsection{Quantum Cover-Leung Region}
The rate region below is based on a coding scheme
where one transmitter decodes the other's message in full.\\
Define the (Quantum Cover-Leung) rate region $\mathcal{R}_{\text{QCL}}(\mathcal{N})$ as follows,
\begin{align}
    \mathcal{R}_{\text{QCL}}(\mathcal{N}) = 
    \bigcup_{\begin{array}{c}
        p_Up_{X_1|U}p_{X_2|U},\theta_{A_1}^{x_1} \otimes \varphi_{A_2}^{x_2}, \Gamma_{B\to \bar{B}Z_1Z_2}
    \end{array}}
    \left\{ 
    \begin{array}{rl}
    (R_1 ,R_2):
    R_1 &\leq I(X_1;Z_2|UX_2)  \\
    R_2 &\leq I(X_2;Z_1|UX_1)  \\
    R_1+R_2 &\leq  I(X_1X_2;\bar{B}Z_1Z_2)_\omega 
    \end{array}
    \right\} \,.
    \label{equation:quantum_cover_leung_region}
\end{align}
The union is taken over the set of all classical auxiliary variables $(U,X_1, X_2) \sim p_{U}p_{X_1|U}p_{X_2|U}$, product state collections $\{ \theta_{A_1}^{x_1} \otimes \varphi_{A_2}^{x_2} \}$ and quantum instruments $\Gamma_{B\to \bar{B}Z_1Z_2}$.
Note that the classical variables $X_1\Cbar U \Cbar X_2$ form a Markov chain. 
Given such auxiliary variables, states, and an instrument, the state is
\begin{align}
    \omega_{\bar{B}Z_1Z_2}^{ux_1x_2} =
    \left(\Gamma_{B\to \bar{B}Z_1Z_2} \circ \mathcal{N}_{A_1A_2\to B}\right)(\theta_{A_1}^{x_1} \otimes \varphi_{A_2}^{x_2}) \,.
\end{align}

In our coding scheme,
the transmitters employ a coding strategy in which each decodes the other's message using the feedback and leverages this information to encode cooperatively (see Section~\ref{appendix:quantum_cover_leung_proof_outline}). % 
The random variable $U$ represents the information known to both transmitters, and $X_1$ and $X_2$ select Alice~1 and Alice~2's quantum states $\theta_{A_1}^{x_1}$ and $\varphi_{A_2}^{x_2}$, respectively, which are transmitted through the channel.

\begin{theorem}[Quantum Cover-Leung bound]
\label{theorem:quantum_cover_leung}
The capacity of the quantum MAC 
$\mathcal{N}_{A_1A_2\to B}$
with classical feedback satisfies
\begin{align}
   \mathcal{C}_{\text{fb}}(\mathcal{N}) \supseteq \mathcal{R}_{\text{QCL}}(\mathcal{N}) \,.
\end{align}
\end{theorem}
To show achievability, we modify the classical scheme of superposition block Markov coding with backward decoding  % 
to the quantum setting. 
Here, superposition refers to the layering of message encoding.
We apply the % 
quantum multiparty packing lemma \cite{Ding:20p} for decoding at the transmitters and the receiver. The receiver's measurement recovers both messages from the same quantum state, preventing state collapse.
The proof outline is provided in Section~\ref{appendix:quantum_cover_leung_proof_outline}.
The classical Cover-Leung rate region \cite{CoverLeung:81p} is obtained as a special case of Theorem~\ref{theorem:quantum_cover_leung}.

\begin{remark}[Full Decode Limitation]
The quantum measurement generates noisy feedback. With noisy feedback, the approach above is overly restrictive, as it forces each encoder to recover the other's entire message through a degraded link.
The following example demonstrates the limitations of this approach.
\begin{example}
\label{example:suboptimal_mac}
In this example, we consider a classical-quantum MAC that demonstrates the limitation of the Cover-Leung coding approach, where Alice~1 and Alice~2  fully decode each other's messages.
We use the example by Fuchs \cite{Fuchs:02b} showing that the accessible information can be strictly lower than the Holevo information, for a single-user channel.

Consider the special case of a MAC
$\mathcal{N}_{X_1 X_2\to B}$ where one input is degenerate (say, $\abs{\mathcal{X}_1}=1$). Specifically, suppose
$\mathcal{N}_{X_1 X_2\to B}(x_1,x_2)\equiv\mathcal{N}^{(2)}_{X_2\to B}(x_2)$ for $(x_1,x_2)\in \{0,1\}^2$, where
\begin{align}
    &\mathcal{N}^{(2)}_{X_2\to B}(0)=\ketbra{0}, \,  \mathcal{N}^{(2)}_{X_2\to B}(1)=\ketbra{+}
    \,.
\end{align}
Note that 
since Alice~1’s input is degenerate, her information rate is $R_1=0$.
Whereas, the capacity of the single-user classical-quantum channel $\mathcal{N}^{(2)}_{X_2\to B}$ is given by the Holevo information $\chi(\mathcal{N}^{(2)})\approx 0.6$ (see \cite[p. 13]{Fuchs:02b}).
Furthermore, classical feedback does not increase the capacity of a quantum single-user channel \cite{Bowen:05p}.
Therefore, the capacity region, either with or without classical feedback, is given by
\begin{align}
\mathcal{C}_{\text{fb}}(\mathcal{N})=
\{ (0,R_2) \,:\; R_2\leq % 
0.6 \} \,.
\end{align}
In our Quantum Cover-Leung region~\eqref{equation:quantum_cover_leung_region}, Alice~2's rate is bounded by  
\begin{align}
     R_2 &\leq I(X_2;Z_1|UX_1)
\end{align}
where $Z_1$ is obtained as a measurement outcome, corresponding to the quantum instrument $\Gamma_{B\to \bar{B}Z_1Z_2}$.
Therefore,
the achievable rate cannot exceed the accessible information % 
(see Section~\ref{section:definitions_channel_model}), i.e.,
\begin{align}
    R_2 &\leq I_\text{acc}(\mathcal{M}_{X_2\to B})  \approx 0.399 
\end{align}
where the last equality is based on the result due to Fuchs
~\cite[p. 13]{Fuchs:02b}.
Therefore, the rate pair $(0, \, 0.6)$ is not achievable using our Quantum Cover-Leung scheme.
The reason for this gap is that our coding scheme requires Alice~1 to decode Alice's~2 message. 
 In the next section, we present an improved region, based on a partial-decode coding scheme that
 avoids this bottleneck. 
 \end{example}
\end{remark}

\subsection{Partial Decode Achievable Region}
\label{subsection:achievable_region}
Define the partial decode (PD) rate region $\mathcal{R}_{\text{PD}}(\mathcal{N})$ as follows,
\begin{align}
    &\mathcal{R}_{\text{PD}}(\mathcal{N}) \equiv
    \bigcup_{\begin{array}{c}
        p_{U}p_{V_1X_1|U}p_{V_2X_2|U} \\
        \theta_{A_1}^{x_1} \otimes \varphi_{A_2}^{x_2} \\
        \Gamma_{B\to \bar{B}Z_1Z_2}
    \end{array}} 
    &\left\{ 
    \begin{array}{rl}
    (R_1 ,R_2):
    R_1 &\leq I(X_1;\bar{B}Z_1Z_2|UV_1X_2)_\omega + I(V_1;Z_2|UX_2)  \\
    R_2 &\leq I(X_2;\bar{B}Z_1Z_2|UV_2X_1)_\omega  + I(V_2;Z_1|UX_1)  \\
    R_1+R_2 &\leq I(V_1;Z_2|X_2U)  +I(V_2;Z_1|X_1U) +  I(X_1X_2;\bar{B}Z_1Z_2|UV_1V_2)_\omega  \\
    R_1+R_2 &\leq I(X_1X_2;\bar{B}Z_1Z_2)_\omega 
    \end{array}
    \right\}
    \label{Equation:Partial_Decode_Achievable_Region}
\end{align}
where the union is taken over the set of all classical auxiliary variables $(U,V_1,X_1,V_2,X_2) \sim p_{U}p_{V_1X_1|U}p_{V_2X_2|U}$, 
state collections $\{ \theta_{A_1}^{x_1} \otimes \varphi_{A_2}^{x_2} \}$ and quantum instruments $\Gamma_{B\to \bar{B}Z_1Z_2}$, where $Z_1$ and $Z_2$ are the measurement outcomes that are sent to Transmitters 1 and 2, respectively and $\bar{B}$ is the post-measurement system at the receiver. 
Note that the classical variables $V_1X_1\Cbar U \Cbar V_2X_2$ form a Markov chain. 
Given such auxiliary variables, states, and an instrument, the output state is
\begin{align}
    \omega_{\bar{B}Z_1Z_2}^{uv_1v_2x_1x_2}=\left(\Gamma_{B\to \bar{B}Z_1Z_2} \circ \mathcal{N}_{A_1A_2\to B}\right)(\theta_{A_1}^{x_1} \otimes \varphi_{A_2}^{x_2}) \,.
\end{align}

In our partial decode scheme, the transmitters only decode a part of the other's message. Hence, $V_1$ represents the information sent from Alice~1 to Alice~2, while $V_2$ represents the information sent from Alice~2 to Alice~1. 

\begin{theorem}
\label{theorem:quantum_inner_bound}
The capacity of the quantum MAC 
$\mathcal{N}_{A_1A_2\to B}$
with classical feedback satisfies
\begin{align}
   \mathcal{C}_{\text{fb}}(\mathcal{N}) \supseteq \mathcal{R}_{\text{PD}}(\mathcal{N}) .
\end{align}
\end{theorem}

\begin{remark}[Feedback Measurement Tradeoff]
\label{remark:achievable_region_tradeoff}
The achievable region highlights a fundamental trade-off due to the feedback measurement; extracting significant information from the state may result in a collapse ($B$ to $\bar{B}$) that could eliminate the quantum advantage for decoding, while avoiding collapse by choosing the measurement $\Gamma=\text{id}_{B\to \bar{B}}$, results in the same rates as without feedback (see Remark~\ref{remark:communication_without_feedback}).
\end{remark}

\begin{remark}[Achievable Regions Comparison]
We observe that  the Partial Decode achievable region is strictly larger than
 the Quantum Cover-Leung achievable region in general.
We first note that $\mathcal{R}_{\text{QCL}}(\mathcal{N}) \subseteq \mathcal{R}_{\text{PD}}(\mathcal{N})$, since the choice of $V_1=X_1$ and $V_2=X_2$ corresponds to the case in which both transmitters fully decode the other’s message (c.f. \eqref{equation:quantum_cover_leung_region} and \eqref{Equation:Partial_Decode_Achievable_Region}).  
Nevertheless, we now % 
show that this inclusion can be strict, i.e.,
there exists a quantum MAC $\mathcal{N}$ such that
\begin{align}
    \mathcal{R}_{\text{QCL}}(\mathcal{N}) \subsetneqq \mathcal{R}_{\text{PD}}(\mathcal{N}).
\end{align}
Consider the channel  in Example~\ref{example:suboptimal_mac}. 
Choosing $\Gamma=\text{id}_{B\to \bar{B}}$ and $U=V_1=V_2=\emptyset$ corresponds to a coding stategy that does not include either generation of feedback or cooperation. Then, we obtain
\begin{align}
    R_2 &\leq I(X_2;\bar{B}Z_1Z_2)_\omega = I(X_2;B)_\omega \,.
\end{align}
Consequently, the rate pair 
$(0, \,  \chi(\mathcal{N}^{(2)})\approx 0.6)$ 
is achievable using the partial-decode coding scheme, i.e., $(0, \,  \chi(\mathcal{N}^{(2)}))\in \mathcal{R}_{\text{PD}}(\mathcal{N})$.
However, as shown in Example~\ref{example:suboptimal_mac}, this rate pair is not achievable using the Quantum Cover-Leung coding scheme, i.e., $(0, \,  \chi(\mathcal{N}^{(2)}))\notin \mathcal{R}_{\text{QCL}}(\mathcal{N})$. 
Hence, the inclusion is
strict.
\end{remark}

\subsection{The Qubit SWAP Channel}
\label{subsection:quantum_binary_adder_mac}

The qubit SWAP channel, as defined by \cite{KlimovitchWinter:05a}
\begin{align}
    \mathcal{N}(\rho_{A_1 A_2})= \frac{1}{2}\rho_{A_1 A_2} + \frac{1}{2}\text{SWAP }\cdot \rho_{A_1 A_2}\cdot \text{ SWAP}^\dagger
\end{align}
where $\text{SWAP}=\sum \ketbra{ji}{ij}$. 
This channel is also referred to as a random permutation channel, or quantum binary adder MAC.
This channel models a setting where the receiver does not know which transmitter sent each qubit.
This is particularly relevant in an optical setup \cite{Frohlich:13p}.
A similar principle stands behind the classical binary adder MAC \cite{CoverLeung:81p} as well.

\begin{figure}[tb]
\begin{center}
\includegraphics[width=0.3\linewidth]{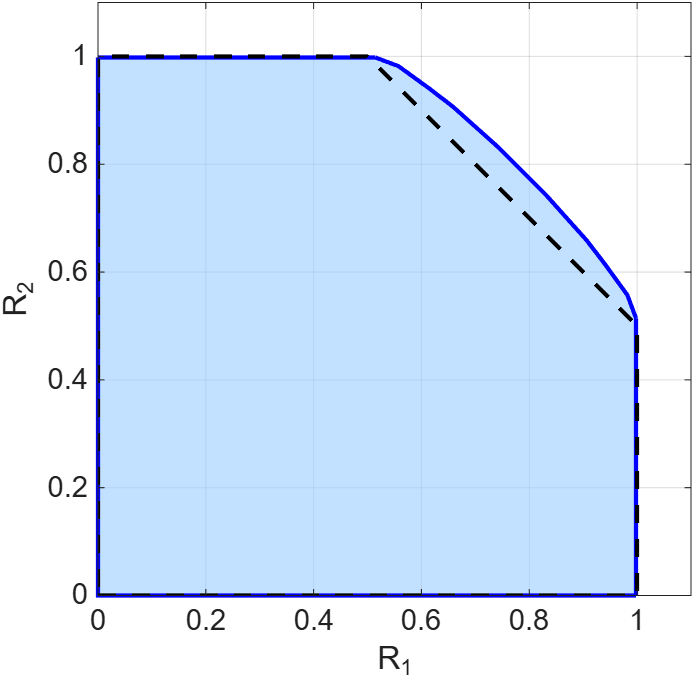}
\end{center}
\caption{Achievable regions for the qubit SWAP channel. The capacity region without feedback is the area 
within the dashed black line. An achievable rate region with classical feedback is indicated by the solid blue line.}
\label{figure:quantum_binary_adder_mac_achievable_region}
\end{figure}

Without feedback, the capacity region is
\cite{KlimovitchWinter:05a}
\begin{align}
\label{equation:binary_adder_capacity_region}
    R_1 \leq 1, \, R_2 \leq 1, \, R_1+R_2 \leq \frac{3}{2} 
    \,.
\end{align}
Based on our quantum Cover-Leung bound in 
Theorem~\ref{theorem:quantum_cover_leung}, we derive an achievable rate region with feedback, as depicted in Fig.~\ref{figure:quantum_binary_adder_mac_achievable_region}. 
To obtain this region, let $U\sim\text{Bernoulli}\left(\frac{1}{2} \right)
$,  $X_1\sim \text{Bernoulli}(\alpha_u)$ and $X_2\sim \text{Bernoulli}(\beta_u)$, with $\alpha_u,\beta_u\in \left[0, \frac{1}{2} \right]$ for $u\in\{0,1\}$, $\theta_{A_1}^{x_1}=\ketbra{x_1}$ and $\varphi_{A_2}^{x_2}=\ketbra{x_2}$. The possible outputs of the channel are
\begin{align}
    \mathcal{N}(\ketbra{00}) = \ketbra{00}_B ,\, 
    \mathcal{N}(\ketbra{11}) = \ketbra{11}_B ,\, \nonumber\\
    \mathcal{N}(\ketbra{10}) = \mathcal{N}(\ketbra{01})  =\frac{1}{2}\ketbra{10}_B +\frac{1}{2}\ketbra{01}_B \,.
\end{align}
We choose the measurement
\begin{align}
    \Gamma_{B\to \bar{B}Z_1Z_2}(\rho_B) = \sum_{y \in \{0,1,2\}}\trace\{{D_y\rho_B}\ketbra{y}_{\bar{B}}\} \otimes \ketbra{y}_{Z_1} \otimes \ketbra{y}_{Z_2}
    \intertext{where}
    \{ \ket{y} \}_{y \in \{0,1,2\}} \, ,D_0 = \ketbra{00}, D_1 = \ketbra{10} +\ketbra{01} , D_2 = \ketbra{11} \,.
\end{align}
If the output is in the support of $D_0$, then we know that the input was $\ket0\otimes \ket0$ with certainty. 
Similarly,  $D_2$ identifies the input as  $\ket1\otimes \ket1$.
Whereas, $D_1$ can be viewed as a confusion subspace, where we have uncertainty regarding the inputs. 

We now develop the rate region in terms of $\alpha_0,\alpha_1,\beta_0,\beta_1$ .
The constraint for $R_1$
\begin{align}
    R_1 &\leq I(X_1;Z_2|UX_2) \nonumber \\
    &= H(X_1|UX_2) -H(X_1|UZ_2X_2)  \nonumber \\
    &= H(X_1|U)\nonumber \\
    &= \frac{1}{2}\left(H(X_1|U=0) + H(X_1|U=1)\right) \nonumber \\
    &= \frac{1}{2}\left(H_2(\alpha_0) + H_2(\alpha_1)\right) 
\end{align}
and in the same manner for $R_2$
\begin{align}
    R_2 \leq \frac{1}{2}\left(H_2(\beta_0) + H_2(\beta_1)\right) \,.
\end{align}
We define $W=\mathbbm{1}_{X_1=X_2}$, such that $\Pr(W=1)=\gamma$.
Expressing $\gamma$ in terms of $\alpha_0,\alpha_1,\beta_0,\beta_1$
\begin{align}
    \gamma = \Pr(W=1)= \Pr(X_1=X_2) = \frac{1}{2}\Big(\alpha_0\beta_0 + (1- \alpha_0)(1-\beta_0)+ \alpha_1\beta_1 + (1- \alpha_1)(1-\beta_1)\Big) \,.
\end{align}
Developing the sum rate constraint:
\begin{align}
    R_1 + R_2 &\leq I(X_1X_2;\bar{B}Z_1Z_2)_\omega \nonumber \\
    &= H(\bar{B}Z_1Z_2)_\omega - H(\bar{B}Z_1Z_2|X_1X_2)_\omega  \nonumber \\
    &= H(\bar{B}Z_1Z_2)_\omega  \nonumber \\
    &= H(W)+H(\bar{B}Z_1Z_2|W)_\omega  \nonumber \\
    &= H_2(\gamma) + \Pr(W=0)H(\bar{B}Z_1Z_2|W=0)_\omega + \Pr(W=1)H(\bar{B}Z_1Z_2|W=1)_\omega  \nonumber \\
    &= H_2(\gamma) + \gamma \cdot H(\bar{B}Z_1Z_2|X_1 = X_2)_\omega \nonumber \\
    &= H_2(\gamma) + \gamma \cdot H_2\left(\frac{P(X_1=1,X_2=1)}{\gamma}\right) \nonumber \\
    &= H_2(\gamma) + \gamma \cdot H_2\left(\frac{\alpha_0\beta_0+\alpha_1\beta_1}{2\gamma}\right) \,.
\end{align}

To conclude, the achievable rate region in terms of $\alpha_0,\alpha_1,\beta_0,\beta_1$
\begin{align}
    R_1 &\leq \frac{1}{2}\left(H_2(\alpha_0) + H_2(\alpha_1)\right) \nonumber \\
    R_2 &\leq  \frac{1}{2}\left(H_2(\beta_0) + H_2(\beta_1)\right) \nonumber \\
    R_1 +R_2 &\leq H_2(\gamma) + \gamma \cdot H_2\left(\frac{\alpha_0\beta_0+\alpha_1\beta_1}{2\gamma}\right)
    \intertext{where}
    &\gamma = \frac{1}{2}\Big(\alpha_0\beta_0 + (1- \alpha_0)(1-\beta_0)+ \alpha_1\beta_1 + (1- \alpha_1)(1-\beta_1)\Big) \,.
\end{align}

\section{Summary and Discussion}
\label{section:summary}
We study the quantum MAC with classical feedback, where feedback is generated through measurement. We then establish an achievable region and give an example, the qubit SWAP channel, where our region with feedback is strictly larger than the capacity region without feedback.

In contrast to classical models, perfect feedback in the quantum setting is impossible due to the no-cloning theorem. Hence, the receiver chooses a measurement to generate feedback, which dictates the noise model. The optimal measurement may depend on the channel. 
Our derivation also yields the classical generalized feedback result \cite[Sec. 11.2]{Kramer:08b}, by replacing
 $\Gamma_{B \to BZ_1Z_2}\circ \mathcal{N}_{A_1 A_2\to B}$ with a general channel from $A_1 A_2$ to $B Z_1 Z_2$, 
 where the channel inputs  and outputs are classical.

Future work includes finding more examples of quantum MACs that benefit from feedback and exploring whether quantum feedback provides additional improvements. A central challenge is to compare classical feedback and entanglement assistance, and determine in which cases each offers greater benefits.

Our findings reveal, for the first time, that classical feedback can expand achievable rates in quantum multi-user communication, opening new directions for the study of hybrid classical-quantum networks.

\section*{Acknowledgments}
The authors thank Gerhard Kramer (TUM)
for useful discussions.
Elina Levi and Uzi Pereg were supported by  ISF, % 
 Grants n. 939/23 and 2691/23,
 DIP % 
 n. 2032991, Ollendorff-Minerva Center % 
 n. 86160946, and  % 
 Helen Diller Quantum Center % 
 n. 	2033613.
Uzi Pereg was also supported by Chaya Chair n. 8776026, and % 
VATAT % 
Program for Quantum Science and Technology % 
n. 86636903. % 

\begin{appendices}
{
\section{A quantum multiparty packing lemma}
\label{appendix:quantum_multiparty_packing_lemma}
In the achievability proof of Theorem~\ref{theorem:quantum_inner_bound} we build upon the quantum multiparty packing lemma \cite{Ding:20p}. We now supply the details for the lemma.

\subsection{Codebook Generation} 
\label{subsection:codebook_generation}
\subsubsection{Multiplex Bayesian network} 
\label{subsubsection:multiplex_bayesian_network}
A multiplex Bayesian network, denoted by $\mathcal{B}=(G,X,\mathcal{M}, \text{ind})$ is used to describe a random codebook structure. It can be interpreted as a mathematical formalization of Markov encoding schemes.
\begin{definition}[Multiplex Bayesian Network]
\label{definition:multiplex_bayesian_network}
A multiplex Bayesian network consists of:
\begin{itemize}
    \item A directed acyclic graph (DAG) $G=(V,E)$ where each vertex $v\in V$ represents a set of codewords that will be generated, conditioned on the codewords of its parents $\text{pa}(v)=\{v'\in V: (v',v)\in E\}$.
    \item A random vector $X$ that consists of random variables $X_v$ with alphabet $\mathcal{X}_v$ for each $v\in V$. This defines the distribution of the codewords to be generated.
    \item A Cartesian product of message sets $\mathcal{M} = \bigtimes\limits_{j\in J}\mathcal{M}_j$, where $J$ denotes the message sets indices. 
    \item A function $\text{ind}:V\to \mathcal{P}(J)$, where $\mathcal{P}(J)$ denotes the power set of $J$, that maps codewords to the message sets it encodes, that satisfies
    \begin{align}
        \text{ind}(v') \subseteq \text{ind}(v) \text{ for } v'\in \text{pa}(v) \nonumber
        \intertext{where}
        \text{pa}(v)=\{v' \in V: (v',v)\in E\}
    \end{align}
    that is, if a parent of $v$, denoted by $v'$, encodes the message set $\mathcal{M}_j$, then $v$ also encodes it. This makes sense, since the codeword $x_v$ depends on $x_{v'}$.
\end{itemize}
\end{definition}

\begin{figure}
\begin{center}
\includegraphics[width=0.3\linewidth]{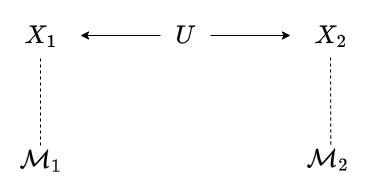}
\end{center}
\caption{MAC Bayesian multiplex network.}
\label{figure:bayesian_multiplex_network_MAC}
\end{figure}

\begin{example}
\label{example:mac}
As a simple example, we give a multiplex Bayesian network used to generate  a random codebook for the 2-user classical-quantum MAC, as illustrated in Fig.~\ref{figure:bayesian_multiplex_network_MAC}. In this scheme, $u$ is generated according to $P_U$, then, $x_1$ encodes a message $m_1\in \mathcal{M}_1$ of Transmitter 1 and is generated according to $P_{X_1|U}$, and $x_2$ encodes a message $m_2\in \mathcal{M}_2$ of Transmitter 2 and is generated according to $P_{X_2|U}$.

We now describe the multiplex Bayesian network; Let $G$ be a graph with $3$ vertices, corresponding to random variables $UX_1X_2 \sim p_{UX_1X_2}$. The graph has edges going from $U$ to $X_1$ and from $U$ to $X_2$. Let $\mathcal{M}_{1}$ and $\mathcal{M}_{2}$ be message sets of Transmitter 1 and Transmitter 2, respectively, where $|\mathcal{M}_{1}| = 2^{nR_1} $ and $|\mathcal{M}_{2}|= 2^{nR_2}$. The function ind maps $U$ to $\emptyset$, $X_1$ to $\{1\}$ and $X_2$ to $\{2\}$, where each mapping is illustrated in Fig.~\ref{figure:bayesian_multiplex_network_MAC} as a dashed line. Then, $X \equiv UX_1X_2$ represents 3 sets of codewords $\{u(\cdot), x_1(\cdot), x_2(\cdot) \}$, $\mathcal{M}=\mathcal{M}_1\times \mathcal{M}_2$ are the message sets and $J=\{1,2\}$ are the message sets indices.
\end{example}

\subsubsection{Codebook generation algorithm} 
\label{subsubsection:codebook_generation_algorithm}
Now, we describe the codebook generation algorithm. The input to the algorithm is a multiplex Bayesian network as defined in Definition~\ref{definition:multiplex_bayesian_network}, and the output is a codebook $C=\{x(m)\in \mathcal{X}\}_{m \in \mathcal{M}}$, where $m$ is an ordered tuple of messages (as $\mathcal{M}$ is a Cartesian product of message sets). We note that each codeword is formally a function over the entire message set $\mathcal{M}$. However, each codeword depends only on the specific messages it encodes, as indicated by line~(5) of the algorithm below.

\fbox{% 
\parbox{0.8\textwidth}{

Algorithm 1: \\
1. \textbf{for} $v\in V$  \textbf{do} \\ 
2. \hspace*{1cm} \textbf{for} $m_v\in \mathcal{M}_{\text{ind}(v)}$  \textbf{do} \\
3. \hspace*{2cm} generate $x_v(m_v)$ according to $P_{X_v|X_{\text{pa}(v)}}(\cdot|x_\text{pa}(m_{\text{pa}(v)}))$ \\
4. \hspace*{2cm} \textbf{for} $m_{\overline{v}}\in \mathcal{M}_{\overline{\text{ind}(v)}}$  \textbf{do} \\
5. \hspace*{3cm} $x_v(m_v, m_{\overline{v}})=x_v(m_v)$ \\
6. \hspace*{2cm} \textbf{end for} \\
7. \hspace*{1cm} \textbf{end for} \\
8. \textbf{end for} 

}
}

The algorithm works as follows;
For every vertex $v\in V$ in $G$, that represent a codeword, go over every ordered tuple of messages it encodes $m_v$. 
Generate a temporary codeword $x_v(m_v)$ according to the conditional distribution of $X_v$ given its parents $X_{\text{pa}(v)}$, where $m_{\text{pa}(v)}$ is a restriction of $m_v$ to $\mathcal{M}_{\text{ind}(\text{pa}(v))}$.
As noted before, each codeword is formally a function over the entire message set $\mathcal{M}$. We set $x_v=x_v(m_v)$, therefore, the codeword does not depend on messages it does not encode.
We note that since $G$ is a DAG, its vertices have a topological order such that for every edge $(v',v)\in E$, $v'$ precedes $v$, hence, line $(3)$ of the algorithm is valid.

In Example~\ref{example:mac}, the algorithm can run as follows:
\begin{enumerate}
    \item $U$: $\text{ind}(u)$ is empty, thus, a single codeword is generated 
    \item $X_1$: $\text{ind}(x_1)=1$, hence $x_1$ is generated according to $P_{X_1|U}$, and depends only on $m_1 \in \mathcal{M}_1$
    \item $X_2$: $\text{ind}(x_2)=2$, hence $x_2$ is generated according to $P_{X_2|U}$, and depends only on $m_2 \in \mathcal{M}_2$.
\end{enumerate}

\begin{definition}[Classical-Quantum output state]
Define
\begin{align}
\label{equation:marginals_density_operator}
    \rho_{XB}^{(\{X_S,B\})}= \sum_{x_{\overline{S}}}p_{X_{\overline{S}}}(x_{\overline{S}})\ketbra{x_{\overline{S}}}_{X_{\overline{S}}} \otimes \rho_{X_S}^{(x_{\overline{S}})}\otimes \rho_{B}^{(x_{\overline{S}})}
\end{align}
where $S \subseteq V, \overline{S}= V \setminus S$.
\end{definition}

\subsection{The quantum hypothesis-testing relative entropy}
Next, we recall a quantum information measure required for our analysis.
The quantum hypothesis-testing relative entropy \cite{Hirche:18t} is defined as:
\begin{align}
    D_{H}^\varepsilon(\rho || \sigma)= \underset{\substack{0\leq \Pi \leq I \\ \trace(\Pi \rho)\geq 1- \varepsilon }}{\max}-\log \trace (\Pi \sigma).
\end{align}

The quantum hypothesis-testing relative entropy quantifies the distinguishability between two quantum states $\rho$ and $\sigma$, corresponding to the null and the alternative hypothesis, respectively. We maximize over all binary-outcome POVMs $\{\Pi, I-\Pi\}$, where $\Pi$ accepts $\rho$ and $I-\Pi$ accepts $\sigma$.
We set the maximal probability of false detection by 
$\varepsilon$, 
and optimize the probability of missed detection of the alternative hypothesis, $\trace (\Pi \sigma)$.
Furthermore, we recall that according to the quantum Stein's lemma~\cite{HiaiPetz:91p}
\begin{align}
    \underset{n \to \infty}{\lim}\frac{1}{n}D_{H}^\varepsilon(\rho^{\otimes n} || \sigma^{\otimes n}) = D(\rho || \sigma) \,.
\end{align}

\subsection{Decoding POVM}
The packing lemma allows us to construct different decoders from the same multiplex Bayesian network. To specify a decoder, we use an ancestral subgraph $H\subseteq G$, consisting of selected vertices $V_H\subseteq V$, together with their parents. Therefore, $v \in V_H$ implies $\text{pa}(v) \subseteq V_H$. We denote the associated random variables by $X_H$, message indices by $J_H$, message set by $\mathcal{M}_H$,  and codebook by $C_H$.

Now we restate the quantum multiparty packing lemma
\begin{lemma} 
[One-shot quantum multiparty packing lemma {\cite[Lem. 2]{Ding:20p}}]
\label{lemma:multiparty_one_shot_packing_lemma} 

Let $\mathcal{B}=(G,X,\mathcal{M}, \text{ind})$ be a multiplex Bayesian network. Run the codebook generation algorithm, Algorithm~1, to obtain a random codebook $C=\{x(m)\in \mathcal{X}\}_{m \in \mathcal{M}}$. Let $H\subseteq G$ be an ancestral subgraph, and $\{\rho_B^{(x_H)}\}_{x_H\in \mathcal{X}_H}$, be a family of quantum states, fix $\varepsilon\in(0,1)$.
Furthermore, consider an index set $D\subseteq J_H$.
Then, there exists a POVM $\{Q_B^{(m_D|m_{\overline{D}})}\}_{m_D\in \mathcal{M}_D}$ for each $m_{\overline{D}}\in \mathcal{M}_{\overline{D}} \equiv \mathcal{M} \setminus  \mathcal{M}_D$, such that for all $(m_D,m_{\overline{D}})\in \mathcal{M}_H$
\begin{align}
    \mathbb{E}_{C_H} \left[ \trace \left[ (I-Q_{B}^{(m_D|m_{\overline{D}})})\rho_{B}^{(x_{H}(m_D,m_{\overline{D}}))} \right] \right] \leq f(|V_H|, \varepsilon) + 4\sum_{ \emptyset \neq L \subseteq D} 2^{(\sum_{\ell \in L}R_\ell)-D_H^\varepsilon(\rho_{X_HB} ||\rho_{X_HB}^{(\{X_{S_L},B\})})}
\end{align}
where  $S_L \equiv \{ v \in V_H | \text{ ind}(v) \cap L \neq \emptyset\}$,
$\omega_{X_HB}= \sum_{x_H\in \mathcal{X}_H} p_{X_H}(x_H)\ketbra{x_H}_{X_H} \otimes \omega_B^{(x_H)}$ , and
$\rho_{X_HB}^{(\{X_S,B\})}$ is defined in \eqref{equation:marginals_density_operator},
with $f(k, \varepsilon) \underset{\varepsilon \to 0}{\longrightarrow} 0$.
\end{lemma}

\subsection{Application to Error Probability Analysis}
To translate the results of the lemma to conditional mutual information, we recall that the conditional mutual information is an asymptotic limit of the hypothesis-testing relative entropy \cite[Eq. (6)]{Ding:20p}:
\begin{align}
\label{equation:HT_relative_entropy}
    &\underset{n \to \infty}{\lim} \frac{1}{n}D_H^{\varepsilon}\left( \rho_{XB}^{\otimes n} || \left(\rho_{XB}^{(\{X_S,B\})}\right)^{\otimes n} \right) \nonumber \\
    & = D\left(\rho_{XB} || \rho_{XB}^{(\{X_S,B\})}\right)\nonumber \\
    &= \sum_{x_{\overline{S}}} p_{X_{\overline{S}}}(x_{\overline{S}}) D\left(\rho_{X_{S}B}^{(x_{\overline{S}})} || \rho_{X_{S}}^{(x_{\overline{S}})} \otimes \rho_{B}^{(x_{\overline{S}})}\right) \nonumber \\
    &= \sum_{x_{\overline{S}}} p_{X_{\overline{S}}}(x_{\overline{S}}) I(X_S;B)_{\rho^{(x_{\overline{S}})}} \nonumber \\
    &= I(X_S;B|X_{\overline{S}})_{\rho} \,.
\end{align}
Hence, for an $n$-fold product state $\rho_{XB}^{\otimes n}$, the right-hand side of \eqref{equation:HT_relative_entropy} tends to zero as $n \to \infty$, provided that
\begin{align}
    \label{equation:mutual_information_bound}
    \sum_{\ell \in L}R_\ell < I(X_{S_L};B|X_{\overline{S_L}})_{\rho} - \delta
\end{align}
where $\overline{S_L}=V_H \setminus S_L$ and $\delta >0$ is arbitrary small.

\section{Proof of Theorem~\ref{theorem:quantum_inner_bound}}

\subsection{Proof outline for Theorem~\ref{theorem:quantum_cover_leung}}
\label{appendix:quantum_cover_leung_proof_outline}
We begin with the proof outline for the case where Alice~1 uses the feedback in order to estimate the (entire) message of Alice~2. Fix a given input ensemble $\left\{p_{U}(u)p_{X_1|U}(x_1|u)p_{X_2|U}(x_2|u) ,\theta_{A_1}^{x_1} \otimes \varphi_{A_2}^{x_2} \right\}$, and a feedback quantum instrument $\Gamma_{B\to \bar{B}\bar{Z}_1\bar{Z}_2Z_1Z_2}$, where $\bar{Z}_1$ and $\bar{Z}_2$ denote local copies of $Z_1$ and $Z_2$ at Bob's.

We use $T$ transmission blocks, each consists of $n$ channel uses, to send a sequence of messages. In the superposition block Markov scheme, Alice~1 and Alice~2 transmit new information in each block, along with old information that helps Bob resolve the remaining uncertainty from the prior block. The old information corresponds to Alice~2's message from the prior block, which Alice~1 recovers using the feedback. The code construction and encoding are given below.

\begin{figure}[tb]
\begin{center}
\includegraphics[width=0.6\linewidth]{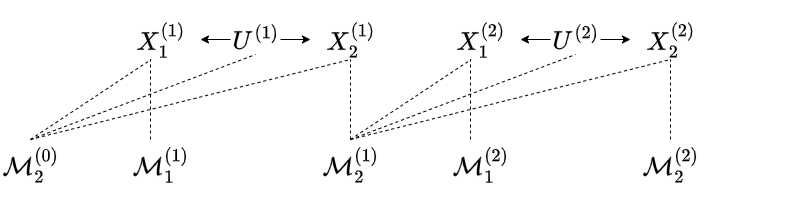}
\end{center}
\caption{Bayesian multiplex network $\mathcal{B}_{\text{cl-fb}}$ generating the codebook $C$ for the quantum MAC with classical feedback, illustrating the Cover–Leung scheme with $T = 2$ blocks.}
\label{figure:bayesian_multiplex_network_MAC_feedback}
\end{figure}

\subsubsection{Classical Code Construction}
To construct the codebook, we use the codebook generation algorithm \cite{Ding:20p}. In block $t \in [1:T]$, Alice $k$ selects a message from $\mathcal{M}_k^{(t)}$, where $\mathcal{M}_2^{(0)}=\mathcal{M}_2^{(T)}=\{1\}$ by convention. The codewords are selected at random, according to the distribution $p_Up_{X_1|U}p_{X_2|U}$, where $\mathbf{u}^{(t)}$ and $\mathbf{x}_k^{(t)}$ encode $m_2^{(t-1)}$ and $m_k^{(t)}$, respectively. 
The codebook structure is illustrated in Fig.~\ref{figure:bayesian_multiplex_network_MAC_feedback} for $T = 2$ blocks.  The codebook is given by
$
    \mathbf{C}=\bigcup_{j=1}^T\left\{ \mathbf{u}(m_2^{(t-1)}),\mathbf{x_1}(m_2^{(t-1)},m_1^{(t)}),\mathbf{x_2}(m_2^{(t-1)},m_2^{(t)}) \right\} ,
$
where we use the short notation $\mathbf{u}\equiv \mathbf{u}^{(t)}$, $\mathbf{x}_k\equiv \mathbf{x}_k^{(t)} \text{ for }k\in \{1,2\}$.

\begin{figure}
\label{figure:encoding_quantum_cover_leung}
\begin{center}
\begin{tabular}{l|ccccc}
{\small Block}				
& $1$ & $2$ & $\cdots$ & $T-1$ & $T$ \\
								
\hline &&&&& \\ 
$X_1$	&
$\mathbf{x_1}(m_1^{(1)}|1)$ &
$\mathbf{x_1}(m_1^{(2)}|\Tilde{m}_2^{(1)})$	&
$\cdots$ &
$\mathbf{x_1}(m_1^{(T-1)}|\Tilde{m}_2^{(T-2)})$ &
$\mathbf{x_1}(m_1^{(T)}|\Tilde{m}_2^{(T-1)})$ 
\\&&&&& \\

$(X_1, Y)$	&
$\Tilde{m}_2^{(1)} \to$ &
$\Tilde{m}_2^{(2)} \to$	&
$\cdots$ &
$\Tilde{m}_2^{(T-1)} \to$ &
$\emptyset$ 
\\&&&&& \\

$X_2$	&
$\mathbf{x_2}(m_2^{(1)}|1)$ &
$\mathbf{x_2}(m_2^{(2)}|m_2^{(1)})$	&
$\cdots$ &
$\mathbf{x_2}(m_2^{(T-1)}|m_2^{(T-2)})$ &
$\mathbf{x_2}(1|m_2^{(T-1)})$ 
\\&&&&& \\

$B$ &
$\hat{m}_1^{(1)}$ &
$\leftarrow\hat{m}_1^{(2)}\hat{m}_2^{(1)}$ &
$\cdots$ &
$\leftarrow\hat{m}_1^{(T-1)}\hat{m}_2^{(T-2)}$ & $\leftarrow\hat{m}_1^{(T)}\hat{m}_2^{(T-1)}$
\\
\end{tabular}
\caption{The block index $t\in [1:T]$ is indicated at the top. In the following rows, we have the corresponding elements: 
(1) codeword of Alice~1;
(2) Alice~1 estimates;
(3) codeword of Alice~2;
(4) estimated messages at Bob.
The arrows in the second row indicate that Alice~1 estimates and encodes forward with respect to the block index, while the arrows in the fourth row indicate that Bob decodes backwards.  }

\end{center}
\end{figure}

\subsubsection{Encoding and Feedback}

At the beginning of block $t\in[1:T]$, % 
Alice~1 uses the received feedback to find a unique $\Tilde{m}_2^{(t-1)}$ such that $(\mathbf{u},\mathbf{x}_1,\mathbf{x}_2,\mathbf{z}_1)$ are jointly typical,
using the estimate $\Tilde{m}_2^{(t-2)}$ from the pervious block.
If none or more are found, use an arbitrary estimation. 
Alice~1 encodes $\mathbf{x_1}(m_1^{(t)}|\Tilde{m}_2^{(t-1)})$ and sends 
$\omega_{\mathbf{A_1}}^{\mathbf{x_1}} =\bigotimes\limits_{i=1}^n \theta_{A_1}^{x_{1,i}^{(t)}}$
using $n$ transmissions via the channel. Alice~2 encodes $\mathbf{x_2}(m_2^{(t)}|m_2^{(t-1)})$ and sends 
$\omega_{\mathbf{A_2}}^{\mathbf{x_2}} =\bigotimes\limits_{i=1}^n \varphi_{A_2}^{x_{2,i}^{(t)}}$
using $n$ transmissions via the channel.

At time $i$, Bob receives
$\omega_{B_i}^{(x_{1,i}^{(t)},x_{2,i}^{(t)})} = \mathcal{N}(\theta_{A_1}^{x_{1,i}^{(t)}} \otimes \varphi_{A_2}^{x_{2,i}^{(t)}})$
, performs the measurement
$\Gamma$ % 
, and transmits the outcomes $z_{1,i}^{(t)}$ and $z_{2,i}^{(t)}$ to Alice~1 and Alice~2, respectively, via the feedback links. 

\subsubsection{Backward Decoding}
The message decoding is performed successively backwards after all $T$ blocks are received. We apply Lemma~\ref{lemma:multiparty_one_shot_packing_lemma} (see \cite{Ding:20p}) for $t = T,$ $T-1,$ $\cdots ,$ $1$. By Lemma~\ref{lemma:multiparty_one_shot_packing_lemma}, there exists a POVM % 
\begin{align} \label{equation:quantum_packing_lemma_POVM}
    \left\{Q_{\mathbf{\bar{B}\bar{Z}_1\bar{Z}_2}}^{\left(m_1^{(t)},m_2^{(t-1)}|\hat{m}_2^{(t)}\right)}\right\}_{m_1^{(t)},m_2^{(t-1)}\in \mathcal{M}_1^{(t)}\times \mathcal{M}_2^{(t-1)}}
\end{align}
where $\hat{m}_2^{(t)}$ is the estimation from the previous decoding step. 
The encoding and decoding procedures are described in Fig.~\ref{figure:encoding_quantum_cover_leung}.

\subsubsection{Error analysis}
We denote Alice~$k$'s messages by $m_k^{[T]}\equiv (m_k^{(0)},\ldots,m_k^{(T)})$, and Alice~1's estimation for Alice~2 messages by $\Tilde{m}_2^{[T]}\equiv (\Tilde{m}_2^{(0)},\ldots,\Tilde{m}_2^{(T)})$.
Define the event that Alice~1 estimates erroneously in block $t$ by
$
    \mathcal{E}_1(t)= \{\Tilde{m}_2^{(t-1)} \neq m_2^{(t-1)}\} 
$
, and the event of Bob's erroneously decoding in block $t$ by
$
    \mathcal{E}_2(t)= \{ (\hat{m}_1^{(t)},\hat{m}_2^{(t-1)})\neq (m_1^{(t)},m_2^{(t-1)})\}
$.

The expected probability of error is bounded by
\begin{align}
     \mathbb{E}_{C_H^n} 
     \left[\bar{P}_e(C)\right] &\leq\sum_{t=1}^{T} 
     \Pr(\mathcal{E}_1(t)| \mathcal{E}_1^c(t-1))
     +  \sum_{t=1}^{T}
     \Pr(\mathcal{E}_2(t)| \mathcal{E}_2^c(t+1) \cap \mathcal{E}_1^c(t))
\end{align}
by the weak law of large numbers and the quantum packing lemma, the error probability tends to zero for each block if
\begin{align}
    R_1 &\leq I(X_1;\bar{B}Z_1Z_2|X_2U)_\omega \nonumber\\
    R_2 &\leq I(X_2;Z_1|X_1U)\nonumber\\
    R_1+R_2 &\leq I(X_1X_2;\bar{B}Z_1Z_2)_\omega \,.
\end{align}
By switching roles between Alice~1 and Alice~2, and using time sharing, we obtain our Quantum Cover–Leung region (see Theorem~\ref{theorem:quantum_cover_leung}). This completes the achievability proof outline.

\subsection{Proof of Theorem~\ref{theorem:quantum_inner_bound} }
\label{appendix:quantum_inner_bound_proof}
Consider a quantum MAC $\mathcal{N}_{A_1A_2\to B}$. 
To prove achievability, we combine quantum information-theoretic tools with the classical feedback approach to construct a code for the quantum MAC with feedback.
Specifically, we adapt the classical block Markov coding scheme using a three-layered superposition code with backward decoding to the quantum setting. 

Fix a given input ensemble $\left\{p_{U}(u)p_{V_1X_1|U}(v_1,x_1|u)p_{V_2X_2|U}(v_2,x_2|u) ,\theta_{A_1}^{x_1} \otimes \varphi_{A_2}^{x_2} \right\}$, and a feedback quantum instrument $\Gamma_{B\to \bar{B}\bar{Z}_1\bar{Z}_2Z_1Z_2}$, where $\bar{Z}_1$ and $\bar{Z}_2$ denote local copies of $Z_1$ and $Z_2$ at Bob's. Denote the channel output by
\begin{align}
    \omega_{B}^{uv_1v_2x_1x_2} = \mathcal{N}_{A_1A_2\to B}(\theta_{A_1}^{x_1} \otimes \varphi_{A_2}^{x_2})
\end{align}
and after the feedback 
\begin{align}
    \omega_{\bar{B}\bar{Z}_1\bar{Z}_2Z_1Z_2}^{uv_1v_2x_1x_2} = \Gamma_{B\to \bar{B}\bar{Z}_1\bar{Z}_2Z_1Z_2}(\omega_{B}^{uv_1v_2x_1x_2}) \,.
\end{align}

We use rate splitting. Alice~$k$'s message is divided into two parts, $m_k = (m_k', m_k'')$, for $k \in \{1,2\}$. 
We use $T$ transmission blocks, each consists of $n$ channel uses, to send a sequence of messages.
In the superposition block Markov scheme, Alice~1 and Alice~2 transmit new information in each block, along with old information that helps Bob resolve the remaining uncertainty from the prior block. The old information corresponds to part of Alice~1's and Alice~2’s messages from the prior block, which they recover using the feedback.
The idea behind rate splitting is to allow Alice~1 and Alice~2 to decode only part of the other’s message from the previous block.
The code construction and encoding are given below.

\subsubsection{Classical codebook generation} \label{qmac_c-fb_codebook}
A random coding strategy is used, in which we generate conditionally independent codewords for Alice~1 and Alice~2, denoted by $\mathbf{v_1},\mathbf{x_1}$ and $\mathbf{v_2},\mathbf{x_2}$, respectively, given a base codeword $\mathbf{u}$. The codewords are drawn according to the fixed distribution $p_Up_{V_1X_1|U}p_{V_2X_2|U}$. 

\begin{figure}[tb]
\begin{center}
\includegraphics[width=1\linewidth]{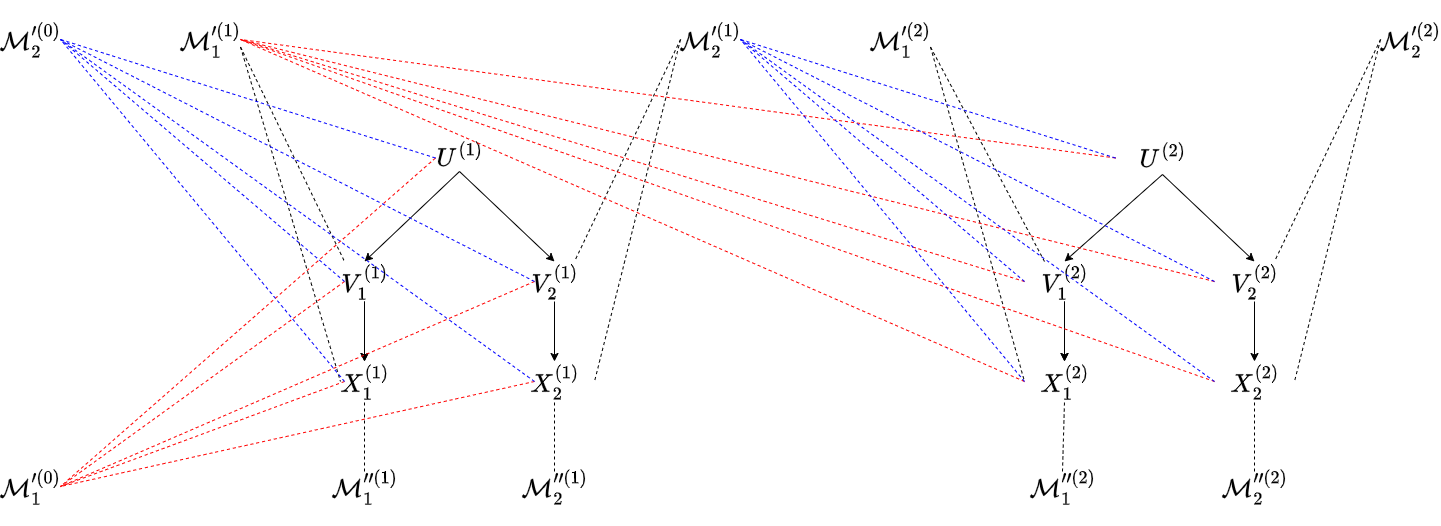}
\end{center}
\caption{Bayesian multiplex network $\mathcal{B}$ for generation of the codebook $C$ for the quantum MAC with feedback rate splitting scheme, with T = 2 blocks.}
\label{figure:bm_rate_splitting_feedback_graph}
\end{figure}

To construct the codebook, we use the codebook generation algorithm described in Sec.~\ref{subsubsection:codebook_generation_algorithm}, Algorithm~1. First, we specify our the corresponding multiplex Bayesian network. Let $G$ be a graph with $5T$ vertices, where $T$ is the total blocks number, random variables $U^{(t)}V_1^{(t)}V_2^{(t)}X_1^{(t)}X_2^{(t)} \sim p_{U}p_{V_1X_1|U}p_{V_2X_2|U}$, where the superscript $(t)$ denotes a single block $t$. The graph has edges going from $U^{(t)}$ to $V_1^{(t)}$ and to $V_2^{(t)}$, from $V_1^{(t)}$ to $X_1^{(t)}$ and from $V_2^{(t)}$ to $X_2^{(t)}$ for all $t$'s and no edges going across blocks with different $t$'s. Let $\mathcal{M}_1'^{(t)},\mathcal{M}_1''^{(t)},\mathcal{M}_2'^{(t)},\mathcal{M}_2''^{(t)}$ be message sets, where $\mathcal{M}_1^{(t)} = \mathcal{M}_1'^{(t)} \times \mathcal{M}_1''^{(t)}$, and  $\mathcal{M}_2^{(t)} = \mathcal{M}_2'^{(t)} \times \mathcal{M}_2''^{(t)}$, and where
$|\mathcal{M}_1'^{(0)}|=|\mathcal{M}_2'^{(0)}|=|\mathcal{M}_1'^{(T)}|=|\mathcal{M}_1''^{(T)}|=|\mathcal{M}_2'^{(T)}|=|\mathcal{M}_2''^{(T)}|=1$ and $|\mathcal{M}_1'^{(t)}| = 2^{nR'_1} $, $|\mathcal{M}_2'^{(t)}|= 2^{nR'_2}$, $|\mathcal{M}_1''^{(t)}| = 2^{nR''_1} $, $|\mathcal{M}_2''^{(t)}|= 2^{nR''_2}$ otherwise. 
The function ind maps:
\begin{itemize}
    \item $U^{(t)}$ to $\mathcal{M}_1'^{(t-1)},\mathcal{M}_2'^{(t-1)}$
    \item $V_1^{(t)}$ to $\mathcal{M}_1'^{(t)},\mathcal{M}_1'^{(t-1)},\mathcal{M}_2'^{(t-1)}$
    \item $V_2^{(t)}$ to $\mathcal{M}_2'^{(t)},\mathcal{M}_1'^{(t-1)},\mathcal{M}_2'^{(t-1)}$
    \item $X_1^{(t)}$ to $\mathcal{M}_1'^{(t)},\mathcal{M}_1''^{(t)},\mathcal{M}_1'^{(t-1)},\mathcal{M}_2'^{(t-1)}$
    \item $X_2^{(t)}$ to $\mathcal{M}_2'^{(t)},\mathcal{M}_2''^{(t)},\mathcal{M}_1'^{(t-1)},\mathcal{M}_2'^{(t-1)}$
\end{itemize} Then, $X \equiv U^{[T]}V_1^{[T]}V_2^{[T]}X_1^{[T]}X_2^{[T]}$, where $U^{[T]}=(U^{(0)},\ldots,U^{(T)})$, the message sets $\mathcal{M}_1^{[T]}=\bigtimes\limits_{j=0}^T \mathcal{M}_1'^{(t)}\times \mathcal{M}_1''^{(t)}$ and $\mathcal{M}_2^{[T]}=\bigtimes\limits_{j=0}^T \mathcal{M}_2'^{(t)}\times \mathcal{M}_2''^{(t)}$, and $\mathcal{M}=\mathcal{M}_1^{[T]}\times \mathcal{M}_2^{[T]}$. $\mathcal{B}_{\text{cl-fb}}=(G, \mathbf{X}, \mathcal{M}, \text{ind})$ is a multiplex Bayesian network, we use the bold notation $\mathbf{X}$ to denote codewords of length $n$, where $\mathbf{X}=  \mathbf{U}^{[T]}\mathbf{V}_1^{[T]}\mathbf{V}_2^{[T]}\mathbf{X}_1^{[T]}\mathbf{X}_2^{[T]}$. See Fig.~\ref{figure:bm_rate_splitting_feedback_graph} for a visualization when $T=2$. Now, run the codebook generation algorithm with $\mathcal{B}_{\text{cl-fb}}$ to get a random codebook
\begin{align}
   \mathbf{C}=\bigcup_{j=1}^T\Biggl\{ \mathbf{u}(m_1'^{(t-1)},m_2'^{(t-1)}),\mathbf{v_1}(m_1'^{(t)}|m_1'^{(t-1)},m_2'^{(t-1)}),\mathbf{v_2}(m_2'^{(t)}|m_1'^{(t-1)},m_2'^{(t-1)}) \nonumber \\
   ,\mathbf{x_1}(m_1'^{(t)},m_1''^{(t)}|m_1'^{(t-1)},m_2'^{(t-1)}),\mathbf{x_2}(m_2'^{(t)},m_2''^{(t)}|m_1'^{(t-1)},m_2'^{(t-1)}) \Biggr\} \,.
\end{align}
Then, the codebook is revealed to all parties. For simplicity, we use the short notation $\mathbf{u}\equiv \mathbf{u}^{(t)}$, $\mathbf{v}_k\equiv \mathbf{v}_k^{(t)},\mathbf{x}_k\equiv \mathbf{x}_k^{(t)} \text{ for }k\in \{1,2\}$.

\subsubsection{Encoding}
At the beginning of block $t\in[1:T]$, based on the received feedback, Alice~1 finds a unique $\Tilde{m}_2'^{(t-1)}$ such that 
\begin{align}
     \Biggl(\mathbf{u}^{(t-1)}(m_1'^{(t-2)},\hat{m}_2'^{(t-2)}),\mathbf{v}_1^{(t-1)}(m_1'^{(t-1)}|m_1'^{(t-2)},{\hat{m}}_2'^{(t-2)}),\mathbf{x}_1^{(t-1)}(m_1'^{(t-1)},m_1''^{(t-1)}|m_1'^{(t-2)},\hat{m}_2'^{(t-2)}), \nonumber\\
     \mathbf{v}_2^{(t-1)}(\Tilde{m}_2'^{(t-1)}|m_1'^{(t-2)},\hat{m}_2'^{(t-2)}), \mathbf{z}_1^{(t-1)})\Biggr) \in
     \mathcal{T}_{\delta}^{(n)}(UV_1X_1V_2Z_1)
\end{align}
using the estimate $\hat{m}_2'^{(t-2)}$ from the pervious block.  If none or more are found, use an arbitrary estimation. Alice~1 encodes the pair $(m_1'^{(t)}, m_2'^{(t)})$ using a base codeword $\mathbf{u}^{(t)}$. Next, she encodes the first part of her current message, $m_1'^{(t)}$, into a codeword $\mathbf{v}_1^{(t)}$ conditioned on $\mathbf{u}^{(t)}$, and then the second part, $m_1''^{(t)}$, into a codeword $\mathbf{x}_1^{(t)}$ conditioned on $\mathbf{v}_1^{(t)}$.
She prepares the state 
\begin{align}
    \omega_{\mathbf{A_1}}^{\mathbf{x_1}} =\bigotimes\limits_{i=1}^n \theta_{A_1}^{x_{1,i}^{(t)}}
\end{align}
and sends it using $n$ transmissions via the channel. \\
Alice~2 follows an analogous encoding procedure. This results in the channel output
\begin{align}
    \omega_{B_i}^{(x_{1,i}^{(t)},x_{2,i}^{(t)})} =  \mathcal{N}_{A_1A_2 \to B}\left(\theta_{A_1}^{x_{1,i}^{(t)}} \otimes \varphi_{A_2}^{x_{2,i}^{(t)}}\right) \,.
\end{align}

\subsubsection{Feedback generation}
At time $i$, Bob applies the feedback quantum instrument $\Gamma_{B\to \bar{B}\bar{Z}_1\bar{Z}_2Z_1Z_2}$ to the channel output
\begin{align}
    \omega_{\bar{B}_i\bar{Z}_1\bar{Z}_2Z_1Z_2}^{(x_{1,i}^{(t)},x_{2,i}^{(t)})} = \Gamma_{B\ \to \bar{B}\bar{Z}_1\bar{Z}_2Z_1Z_2}(\omega_{B_i}^{(x_{1,i}^{(t)},x_{2,i}^{(t)})})
 \end{align}
where the system $\bar{B}$ denotes the post-measurement system, and $\bar{Z}_1$ and $\bar{Z}_2$ are local copies of the classical $Z_1$ and $Z_2$ preserved for the decoding process.
We denote the classical measurement outcomes $Z_1$ and $Z_2$ by $z_{1,i}^{(t)}$ and $z_{2,i}^{(t)}$; these are transmitted via the feedback links to Alice~1 and Alice~2, respectively.

\subsubsection{Backward Decoding}
\begin{figure}[tb]
\begin{center}
\includegraphics[width=0.7\linewidth]{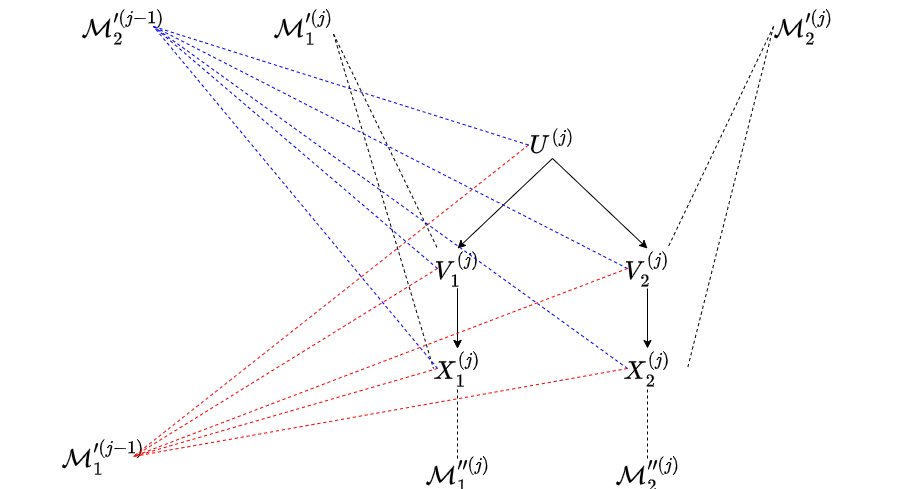}
\end{center}
\caption{An ancestral subgraph of $\mathcal{B}$ in Fig.~\ref{figure:bm_rate_splitting_feedback_graph} for backward decoding of the quantum MAC with feedback.}
\label{figure:bm_rate_splitting_feedback_block_graph}
\end{figure}

The message decoding is performed successively backwards after all $T$ blocks are received. We apply Lemma~\ref{lemma:multiparty_one_shot_packing_lemma} (see \cite{Ding:20p}) for $t = T,$ $T-1,$ $\cdots ,$ $1$ with the multiplex Bayesian network $\mathcal{B}_{\text{cl-fb}}$, the ancestral subgraph $H$ containing vertices $U^{(t)},V_1^{(t)}, V_2^{(t)},X_1^{(t)},X_2^{(t)}$ and $J_H=\{m_1'^{(t-1)},m_2'^{(t-1)},m_1'^{(t)},m_2'^{(t)},m_1''^{(t)},m_2''^{(t)}\}$, as visualized in Fig.~\ref{figure:bm_rate_splitting_feedback_block_graph}, $D=\{m_1'^{(t-1)},m_2'^{(t-1)},m_1''^{(t)},m_2''^{(t)}\}$, $\varepsilon(n)=\frac{1}{n}$ and quantum states
\begin{align}
    \left\{\omega_{\mathbf{\bar{B}}^{(t)}\mathbf{\bar{Z}}_1^{(t)}\mathbf{\bar{Z}}_2^{(t)}}^{\left(\mathbf{x_1}(m_1'^{(t)},m_1''^{(t)}|m_1'^{(t-1)},\hat{m}_2'^{(t-1)}),\mathbf{x_2}(m_2'^{(t)},m_2''^{(t)}|m_1'^{(t-1)},\hat{m}_2'^{(t-1)})\right)}\right\}_{\mathbf{x_1}\in \mathcal{X}_1^n,\mathbf{x_2}\in \mathcal{X}_2^n} \,.
\end{align}
We denote the POVM from the lemma by
\begin{align}
    \left\{Q_{\mathbf{\bar{B}\bar{Z}_1\bar{Z}_2}}^{\left(m_1'^{(t-1)},m_2'^{(t-1)},m_1''^{(t)},m_2''^{(t)}|\hat{m}_1'^{(t)},\hat{m}_2'^{(t)}\right)}\right\}_{m_1'^{(t-1)},m_2'^{(t-1)},m_1''^{(t)},m_2''^{(t)}\in \mathcal{M}_1'^{(t-1)}\times \mathcal{M}_2'^{(t-1)} \times \mathcal{M}_1''^{(t)}\times \mathcal{M}_2''^{(t)}}
\end{align}
where $\hat{m}_1'^{(t)},\hat{m}_2'^{(t)}$ is the estimation from the previous decoding step, and we obtain the estimate $(m_1'^{(t-1)},m_2'^{(t-1)},m_1''^{(t)},m_2''^{(t)})$ from measurement.

\subsubsection{Error Analysis}
We denote Alice~$k$'s messages by $m_k^{[T]}\equiv (m_k^{(0)},\ldots,m_k^{(T)})$. Furthermore, we denote Alice~1's estimation for part of Alice~2 messages by $\Tilde{m}_2'^{[T]}\equiv (\Tilde{m}_2'^{(0)},\ldots,\Tilde{m}_2'^{(T)})$, and Alice~2's estimation for part of Alice~1 messages by $\Tilde{m}_1'^{[T]}\equiv (\Tilde{m}_1'^{(0)},\ldots,\Tilde{m}_1'^{(T)})$. \\
The expected probability of error is thus
\begin{align}
    \mathbb{E}_{C_H^n}\left[P_e^{(n)}(C)\right] & =
    \Pr\Bigl((\hat{m}_1'^{[T]},\hat{m}_1''^{[T]},\hat{m}_2'^{[T]},\hat{m}_2''^{[T]})\neq (m_1'^{[T]},m_1''^{[T]},m_2'^{[T]},m_2''^{[T]})\Bigr) \nonumber \\ 
    &\leq \Pr\left((\hat{m}_1'^{[T]},\hat{m}_1''^{[T]},\hat{m}_2'^{[T]},\hat{m}_2''^{[T]})\neq (m_1'^{[T]},m_1''^{[T]},m_2'^{[T]},m_2''^{[T]})\right)\cup \left( \Tilde{m}_1'^{[T]} \neq m_1'^{[T]} \right)\cup \left( \Tilde{m}_2'^{[T]} \neq m_2'^{[T]} \right) \nonumber\\
    &\leq \Pr\Bigl(\Tilde{m}_1'^{[T]} \neq m_1'^{[T]}\Bigr) + \Pr\Bigl(\Tilde{m}_2'^{[T]} \neq m_2'^{[T]}\Bigr)  \nonumber\\
    &+ \Pr\Bigl((\hat{m}_1'^{[T]},\hat{m}_1''^{[T]},\hat{m}_2'^{[T]},\hat{m}_2''^{[T]})\neq (m_1'^{[T]},m_1''^{[T]},m_2'^{[T]},m_2''^{[T]})|\Tilde{m}_1'^{[T]} = m_1'^{[T]},\Tilde{m}_2'^{[T]} =m_2'^{[T]} \Bigr)  \,.
    \label{equation:error_probability}
\end{align}

Consider the first term, corresponding to Alice~1's estimation. Based on the union of events bound,
\begin{align}
    \Pr\Bigl(\Tilde{m}_1'^{[T]} \neq m_1'^{[T]}\Bigr)
    &\leq \sum_{j=1}^{T}\Pr\left(\Tilde{m}_1'^{(t)} \neq m_1'^{(t)}| \Tilde{m}_1'^{[t-1]} = m_1'^{[t-1]}\right).
\end{align}
Each summand tends to zero by the weak law of large numbers and the packing lemma if
\begin{align}
    R_1' < I(V_1;Z_2X_2|U) -\varepsilon_1(\delta ) =  I(V_1;Z_2|X_2U) + I(V_1;X_2|U) -\varepsilon_1(\delta ) = I(V_1;Z_2|X_2U) -\varepsilon_1(\delta )
\end{align}
where $\varepsilon_1(\delta )=2\delta H(V_1|U)$. \\
In the same manner, the second error term in Eq.~\eqref{equation:error_probability}, 
corresponding to Alice~1's estimation, vanishes if
\begin{align}
    R_2' < I(V_2;Z_1|X_1U) - \varepsilon_2(\delta )
\end{align}
where $\varepsilon_2(\delta )=2\delta H(V_2|U)$.

The last term in Eq.~\eqref{equation:error_probability}, is the probability that Bob has a decoding error conditioned on correct estimations of Alice~1 and Alice~2:
\begin{align}
    &\Pr\Bigl((\hat{m}_1'^{[T]},\hat{m}_1''^{[T]},\hat{m}_2'^{[T]},\hat{m}_2''^{[T]})\neq (m_1'^{[T]},m_1''^{[T]},m_2'^{[T]},m_2''^{[T]})|\Tilde{m}_1'^{[T]} = m_1'^{[T]},\Tilde{m}_2'^{[T]} =m_2'^{[T]} \Bigr)       \nonumber \\ 
    &\leq \sum_{j=1}^{T}\Pr\Bigl((\hat{m}_1'^{(t-1)},\hat{m}_1''^{(t)},\hat{m}_2'^{(t-1)},\hat{m}_2''^{(t)})\neq (m_1'^{(t-1)},m_1''^{(t)},m_2'^{(t-1)},m_2''^{(t)}) \Big| \nonumber \\
    &\hspace*{7cm} \Tilde{m}_1'^{[T]} = m_1'^{[T]},\Tilde{m}_2'^{[T]} =m_2'^{[T]}, \hat{m}_1'^{(t)} = m_1'^{(t)}, \hat{m}_2'^{(t)} = m_2'^{(t)}\Bigr) \nonumber \\
    &=\sum_{j=1}^{T}\mathbb{E}_{C_H^n}\Biggl[\trace\Biggl[ \left(I-Q_{\mathbf{\bar{B}\bar{Z}_1\bar{Z}_2}}^{\left(m_1'^{(t-1)},m_2'^{(t-1)},m_1''^{(t)},m_2''^{(t)}|\hat{m}_1'^{(t)},\hat{m}_2'^{(t)}\right)}\right)  \nonumber \\
    &\hspace*{7cm} \omega_{\mathbf{\bar{B}}^{(t)}\mathbf{\bar{Z}}_1^{(t)}\mathbf{\bar{Z}}_2^{(t)}}^{\left(\mathbf{x_1}(m_1'^{(t)},m_1''^{(t)},m_1'^{(t-1)},\hat{m}_2'^{(t-1)}),\mathbf{x_2}(m_2'^{(t)},m_2''^{(t)},m_1'^{(t-1)},\hat{m}_2'^{(t-1)})\right)}\Biggr]\Biggr].
\end{align}
By the quantum multiparty packing lemma, (Lemma \ref{lemma:multiparty_one_shot_packing_lemma}), each summand is
\begin{align}
    & f(5,\varepsilon(n))+4 \times 2^{n(\sum_{\ell \in L}R_\ell)-D_H^{\varepsilon(n)}(\omega_{\mathbf{U}\mathbf{V_1}\mathbf{X_1}\mathbf{V_2}\mathbf{X_2}\mathbf{\bar{B}\bar{Z}_1\bar{Z}_2}} ||\omega_{\mathbf{U}\mathbf{V_1}\mathbf{X_1}\mathbf{V_2}\mathbf{X_2}\mathbf{\bar{B}\bar{Z}_1\bar{Z}_2}}^{(\{\mathbf{X_{S_L}},\mathbf{\bar{B}\bar{Z}_1\bar{Z}_2}\})})} = \nonumber \\
    &f(5,\varepsilon(n))+4 \times 2^{n(\sum_{\ell \in L}R_\ell)-D_H^{\varepsilon(n)}\left(\left(\omega_{UX_1V_2X_2\bar{B}\bar{Z}_1\bar{Z}_2}\right)^{\otimes n} ||\left(\omega_{UX_1V_2X_2\bar{B}\bar{Z}_1\bar{Z}_2}^{(\{X_{S_L},\bar{B}\bar{Z}_1\bar{Z}_2\})}\right)^{\otimes n}\right)}.
\end{align}
By \eqref{equation:mutual_information_bound}, the error probability vanishes asymptotically if the following conditions are met:
\begin{align}
    R_1' &< I(UV_1X_1V_2X_2;\bar{B}\bar{Z}_1\bar{Z}_2)_\omega - \varepsilon \label{r1}\\
    R_2' &< I(UV_1X_1V_2X_2;\bar{B}\bar{Z}_1\bar{Z}_2)_\omega - \varepsilon \label{r2}\\   
    R_1'' &< I(X_1;\bar{B}\bar{Z}_1\bar{Z}_2|UV_1V_2X_2)_\omega - \varepsilon \\
    R_2'' &< I(X_2;\bar{B}\bar{Z}_1\bar{Z}_2|UV_1V_2X_1)_\omega - \varepsilon \\
    R_1' + R_2' &< I(UV_1X_1V_2X_2;\bar{B}\bar{Z}_1\bar{Z}_2)_\omega - \varepsilon \label{r3}\\
    R_1' + R_1'' &< I(UV_1X_1V_2X_2;\bar{B}\bar{Z}_1\bar{Z}_2)_\omega - \varepsilon \label{r4}\\
    R_1' + R_2'' &< I(UV_1X_1V_2X_2;\bar{B}\bar{Z}_1\bar{Z}_2)_\omega - \varepsilon \label{r5}\\
    R_2' + R_1'' &< I(UV_1X_1V_2X_2;\bar{B}\bar{Z}_1\bar{Z}_2)_\omega - \varepsilon \label{r6}\\
    R_2' + R_2'' &< I(UV_1X_1V_2X_2;\bar{B}\bar{Z}_1\bar{Z}_2)_\omega - \varepsilon \label{r7}\\
    R_1''+ R_2'' &< I(X_1X_2;\bar{B}\bar{Z}_1\bar{Z}_2|UV_1V_2)_\omega - \varepsilon\\
    R_1' + R_1'' +R_2' &< I(UV_1X_1V_2X_2;\bar{B}\bar{Z}_1\bar{Z}_2)_\omega - \varepsilon\label{r8}\\
    R_1' + R_2'' +R_2' &< I(UV_1X_1V_2X_2;\bar{B}\bar{Z}_1\bar{Z}_2)_\omega - \varepsilon\label{r9}\\
    R_1' + R_1'' +R_2'' &< I(UV_1X_1V_2X_2;\bar{B}\bar{Z}_1\bar{Z}_2)_\omega - \varepsilon\label{r10}\\
    R_2' + R_1'' +R_2'' &< I(UV_1X_1V_2X_2;\bar{B}\bar{Z}_1\bar{Z}_2)_\omega - \varepsilon\label{r11}\\
    R_1' + R_1'' + R_2' + R_2'' &< I(UV_1X_1V_2X_2;\bar{B}\bar{Z}_1\bar{Z}_2)_\omega - \varepsilon  \,.
    \label{equation:th3_dominating_sum_rate}
\end{align}
Among the 15 conditions above, % 
11 of which are redundant. Specifically, $\eqref{r1}$-$\eqref{r2}$, $\eqref{r3}$-$\eqref{r7}$, $\eqref{r8}$-$\eqref{r11}$, are  dominated by the bound in  \eqref{equation:th3_dominating_sum_rate}. As we replace $\bar{Z}_1$ and $\bar{Z}_2$ by $Z_1$ and $Z_2$, respectively, the remaining conditions are
\begin{align}
    R_1' &< I(V_1;Z_2|UX_2) - \varepsilon_1(\delta)\\
    R_2' &< I(V_2;Z_1|UX_1) - \varepsilon_2(\delta) \nonumber \\
    R_1'' &< I(X_1;\bar{B}Z_1Z_2|UV_1V_2X_2)_\omega - \varepsilon =  I(X_1;\bar{B}Z_1Z_2|UV_1X_2)_\omega - \varepsilon \nonumber \\
    R_2'' &< I(X_2;\bar{B}Z_1Z_2|UV_1V_2X_1)_\omega - \varepsilon = I(X_2;\bar{B}Z_1Z_2|UV_2X_1)_\omega - \varepsilon \nonumber \\
    R_1''+R_2'' &< I(X_1X_2;\bar{B}Z_1Z_2|UV_1V_2)_\omega - \varepsilon \nonumber \\
    R_1+R_2 &< I(UV_1X_1V_2X_2;\bar{B}Z_1Z_2)_\omega - \varepsilon  =I(X_1X_2;\bar{B}Z_1Z_2)_\omega - \varepsilon \,.
\end{align}
By eliminating $R_1',R_1'',R_2',R_2''$, we obtain the following conditions for reliable communication:
\begin{align}
    R_1 & < I(X_1;\bar{B}Z_1Z_2|UV_1X_2)_\omega + I(V_1;Z_2|UX_2) -\varepsilon -\varepsilon_1(\delta) \nonumber \\
    R_2 & < I(X_2;\bar{B}Z_1Z_2|UV_2X_1)_\omega + I(V_2;Z_1|UX_1) -\varepsilon- \varepsilon_2(\delta) \nonumber \\
    R_1+R_2 & < I(V_1;Z_2|X_2U) +I(V_2;Z_1|X_1U)+ I(X_1X_2;\bar{B}Z_1Z_2|UV_1V_2)_\omega  -2\varepsilon -\varepsilon_1(\delta) - \varepsilon_2(\delta) \nonumber \\
    R_1+R_2 & < I(X_1X_2;\bar{B}Z_1Z_2)_\omega -2\varepsilon -\varepsilon_1(\delta) - \varepsilon_2(\delta) \,.
\end{align}

As the average over all codebook ensembles yield a vanishing error, it follows that there exists a deterministic codebook with the same property. 
Achievability now follows by taking the limits of $n\to\infty$,  $\varepsilon \to 0$ and $\delta\to 0$.
This completes the achievability proof. \qed
}
\end{appendices}

\bibliography{references}

\begin{thebibliography}{10}
\providecommand{\url}[1]{#1}
\csname url@samestyle\endcsname
\providecommand{\newblock}{\relax}
\providecommand{\bibinfo}[2]{#2}
\providecommand{\BIBentrySTDinterwordspacing}{\spaceskip=0pt\relax}
\providecommand{\BIBentryALTinterwordstretchfactor}{4}
\providecommand{\BIBentryALTinterwordspacing}{\spaceskip=\fontdimen2\font plus
\BIBentryALTinterwordstretchfactor\fontdimen3\font minus \fontdimen4\font\relax}
\providecommand{\BIBforeignlanguage}[2]{{%
\expandafter\ifx\csname l@#1\endcsname\relax
\typeout{** WARNING: IEEEtran.bst: No hyphenation pattern has been}%
\typeout{** loaded for the language `#1'. Using the pattern for}%
\typeout{** the default language instead.}%
\else
\language=\csname l@#1\endcsname
\fi
#2}}
\providecommand{\BIBdecl}{\relax}
\BIBdecl

\bibitem{Liu:25p}
J.~Liu, T.~Le, T.~Ji, R.~Yu, D.~Farfurnik, G.~Byrd, and D.~Stancil, ``The road to quantum {I}nternet: Progress in quantum network testbeds and major demonstrations,'' \emph{Prog. Quantum Electron.}, vol.~99, p. 100551, 2025.

\bibitem{Cao:22p}
Y.~Cao, Y.~Zhao, Q.~Wang, J.~Zhang, S.~X. Ng, and L.~Hanzo, ``The evolution of quantum key distribution networks: On the road to the qinternet,'' \emph{IEEE Commun. Surv. Tutor.}, vol.~24, no.~2, pp. 839--894, 2022.

\bibitem{Caleffi:24p}
M.~Caleffi, M.~Amoretti, D.~Ferrari, J.~Illiano, A.~Manzalini, and A.~S. Cacciapuoti, ``Distributed quantum computing: a survey,'' \emph{Comput. Netw.}, vol. 254, p. 110672, 2024.

\bibitem{Aslam:23p}
N.~Aslam, H.~Zhou, E.~K. Urbach, M.~J. Turner, R.~L. Walsworth, M.~D. Lukin, and H.~Park, ``Quantum sensors for biomedical applications,'' \emph{Nat. Rev. Phys.}, vol.~5, no.~3, pp. 157--169, 2023.

\bibitem{Vcarapic:20p}
D.~{\v{C}}arapi{\'c} and M.~Maksimovi{\'c}, ``A comparison of {5G} channel coding techniques,'' \emph{Int. J. Electr. Eng. Comput.}, vol.~4, no.~2, pp. 71--82, 2020.

\bibitem{LukensPetersQi:25a}
J.~M. Lukens, N.~A. Peters, and B.~Qi, ``Hybrid classical-quantum communication networks,'' \emph{arXiv preprint arXiv:2502.07298}, 2025.

\bibitem{Pereg:25p}
U.~Pereg, ``Quantum relay channels,'' in \emph{Proc. IEEE Int. Symp. Inf. Theory (ISIT)}, 2025, accepted for publication in IEEE Trans. Inf. Theory.

\bibitem{IlinPereg:25a}
Y.~Ilin and U.~Pereg, ``Relaying quantum information,'' \emph{arXiv preprint arXiv:2507.06770}, 2025, accepted for publication in Proc. 61st Allerton Conf. Commun. Contr. Comput. (Allerton).

\bibitem{Hawellek:25p}
J.~Hawellek, A.~Mohan, H.~Aghaee, and C.~Deppe, ``The interference channel with entangled transmitters,'' in \emph{Proc. IEEE Int. Symp. Inf. Theory (ISIT)}, 2025.

\bibitem{PeregDeppeBoche:21p}
U.~Pereg, C.~Deppe, and H.~Boche, ``Quantum broadcast channels with cooperating decoders: An information-theoretic perspective on quantum repeaters,'' \emph{J. Math. Phys.}, vol.~62, no.~6, 2021.

\bibitem{FawziFerme:24p}
O.~Fawzi and P.~Fermé, ``Broadcast channel coding: Algorithmic aspects and non-signaling assistance,'' \emph{IEEE Trans. Inf. Theory}, vol.~70, no.~11, pp. 7563--7580, 2024.

\bibitem{YaoJafar:25p}
Y.~Yao and S.~A. Jafar, ``Can non-signaling assistance increase the degrees of freedom of a wireless network?'' in \emph{Proc. IEEE Int. Symp. Inf. Theory (ISIT)}, 2025.

\bibitem{Liu24:p}
Y.~Liu, C.~Ouyang, Z.~Ding, and R.~Schober, ``The road to next-generation multiple access: A 50-year tutorial review,'' \emph{Proc. of the IEEE}, vol. 112, no.~9, pp. 1100--1148, 2024.

\bibitem{Leditzky:20p}
F.~Leditzky, M.~A. Alhejji, J.~Levin, and G.~Smith, ``Playing games with multiple access channels,'' \emph{Nature Commun.}, vol.~11, no.~1, pp. 1--5, 2020.

\bibitem{PeregDeppeBoche:25p}
U.~Pereg, C.~Deppe, and H.~Boche, ``The multiple-access channel with entangled transmitters,'' \emph{IEEE Trans. Inf. Theory}, vol.~71, no.~2, pp. 1096--1120, 2025.

\bibitem{FawziFerme:23p}
O.~Fawzi and P.~Ferm{\'e}, ``Multiple-access channel coding with non-signaling correlations,'' \emph{IEEE Trans. Inf. Theory}, 2023.

\bibitem{Winter:01p}
A.~Winter, ``The capacity of the quantum multiple-access channel,'' \emph{IEEE Trans. Inf. Theory}, vol.~47, no.~7, pp. 3059--3065, 2001.

\bibitem{HsiehDevetakWinter:08p}
M.-H. Hsieh, I.~Devetak, and A.~Winter, ``Entanglement-assisted capacity of quantum multiple-access channels,'' \emph{IEEE Trans. Inf. Theory}, vol.~54, no.~7, pp. 3078--3090, 2008.

\bibitem{LiuChenXi:25p}
X.~Liu, F.~Chen, and Z.~Xi, ``Entanglement-assisted private communication over a classical-quantum multiple-access channel,'' \emph{Phys. Rev. A}, vol. 112, no.~1, p. 012618, 2025.

\bibitem{BocheNotzel:14p}
H.~Boche and J.~N\"otzel, ``The classical-quantum multiple access channel with conferencing encoders and with common messages,'' \emph{Quantum Inf. Process.}, vol.~13, no.~12, pp. 2595--2617, 2014.

\bibitem{PeregDeppeBoche:22p}
U.~Pereg, C.~Deppe, and H.~Boche, ``The quantum multiple-access channel with cribbing encoders,'' \emph{IEEE Trans. Inf. Theory}, vol.~68, no.~6, pp. 3965--3988, 2022.

\bibitem{Xu:24p}
P.~Xu, G.~Chen, Z.~Yang, Y.~Li, and S.~Tomasin, ``Multiple access wiretap channel with partial rate-limited feedback,'' \emph{IEEE Trans. Inf. Theory}, vol.~19, pp. 3279--3294, 2024.

\bibitem{Shannon:56p}
C.~Shannon, ``The zero error capacity of a noisy channel,'' \emph{IRE Trans. Inf. Theory}, vol.~2, no.~3, pp. 8--19, 1956.

\bibitem{Kramer:08b}
G.~Kramer, ``Topics in multi-user information theory,'' \emph{Found. Trends Commun. Inf. Theory}, vol.~4, no. 4-5, pp. 265--444, 2008.

\bibitem{CoverPombra:89p}
T.~Cover and S.~Pombra, ``Gaussian feedback capacity,'' \emph{IEEE Trans. Inf. Theory}, vol.~35, no.~1, pp. 37--43, 1989.

\bibitem{Bowen:05p}
G.~Bowen and R.~Nagarajan, ``On feedback and the classical capacity of a noisy quantum channel,'' \emph{IEEE Trans. Inf. Theory}, vol.~51, no.~1, pp. 320--324, 2005.

\bibitem{GaarderWolf:75p}
N.~Gaarder and J.~Wolf, ``The capacity region of a multiple-access discrete memoryless channel can increase with feedback (corresp.),'' \emph{IEEE Trans. Inf. Theory}, vol.~21, no.~1, pp. 100--102, 1975.

\bibitem{CoverLeung:81p}
T.~Cover and C.~Leung, ``An achievable rate region for the multiple-access channel with feedback,'' \emph{IEEE Trans. Inf. Theory}, vol.~27, no.~3, pp. 292--298, 1981.

\bibitem{BrossLapidoth:05p}
S.~Bross and A.~Lapidoth, ``An improved achievable region for the discrete memoryless two-user multiple-access channel with noiseless feedback,'' \emph{IEEE Trans. Inf. Theory}, vol.~51, no.~3, pp. 811--833, 2005.

\bibitem{VenkataramananPradhan:11p}
R.~Venkataramanan and S.~S. Pradhan, ``A new achievable rate region for the multiple-access channel with noiseless feedback,'' \emph{IEEE Trans. Inf. Theory}, vol.~57, no.~12, pp. 8038--8054, 2011.

\bibitem{Kramer:98t}
G.~Kramer, ``Directed information for channels with feedback,'' Ph.D. dissertation, Swiss Federal Institute of Technology, Zürich, 1998.

\bibitem{PermuterWeissmanChen:09p}
H.~H. Permuter, T.~Weissman, and J.~Chen, ``Capacity region of the finite-state multiple-access channel with and without feedback,'' \emph{IEEE Trans. Inf. Theory}, vol.~55, no.~6, pp. 2455--2477, 2009.

\bibitem{Willems:82p}
F.~Willems, ``The feedback capacity region of a class of discrete memoryless multiple access channels (corresp.),'' \emph{IEEE Trans. Inf. Theory}, vol.~28, no.~1, pp. 93--95, 1982.

\bibitem{Ozarow:84p}
L.~Ozarow, ``The capacity of the white {Gaussian} multiple access channel with feedback,'' \emph{IEEE Trans. Inf. Theory}, vol.~30, no.~4, pp. 623--629, 1984.

\bibitem{Carleial:82p}
A.~Carleial, ``Multiple-access channels with different generalized feedback signals,'' \emph{IEEE Trans. Inf. Theory}, vol.~28, no.~6, pp. 841--850, 1982.

\bibitem{LapidothWigger:10p}
A.~Lapidoth and M.~Wigger, ``On the {AWGN} mac with imperfect feedback,'' \emph{IEEE Trans. Inf. Theory}, vol.~56, no.~11, pp. 5432--5476, 2010.

\bibitem{ShavivSteinberg:13p}
D.~Shaviv and Y.~Steinberg, ``On the multiple-access channel with common rate-limited feedback,'' \emph{IEEE Trans. Inf. Theory}, vol.~59, no.~6, pp. 3780--3795, 2013.

\bibitem{OzarowLeung:84p}
L.~Ozarow and S.~Leung-Yan-Cheong, ``An achievable region and outer bound for the {Gaussian} broadcast channel with feedback (corresp.),'' \emph{IEEE Trans. Inf. Theory}, vol.~30, no.~4, pp. 667--671, 1984.

\bibitem{Gastpar:14p}
M.~Gastpar, A.~Lapidoth, Y.~Steinberg, and M.~Wigger, ``Coding schemes and asymptotic capacity for the {Gaussian} broadcast and interference channels with feedback,'' \emph{IEEE Trans. Inf. Theory}, vol.~60, no.~1, pp. 54--71, 2014.

\bibitem{LaiElGamalPoor:08p}
L.~Lai, H.~El~Gamal, and H.~V. Poor, ``The wiretap channel with feedback: Encryption over the channel,'' \emph{IEEE Trans. Inf. Theory}, vol.~54, no.~11, pp. 5059--5067, 2008.

\bibitem{Ardestanizadeh:09p}
E.~Ardestanizadeh, M.~Franceschetti, T.~Javidi, and Y.-H. Kim, ``Wiretap channel with secure rate-limited feedback,'' \emph{IEEE Trans. Inf. Theory}, vol.~55, no.~12, pp. 5353--5361, 2009.

\bibitem{VenkataramananPradhan:13p}
R.~Venkataramanan and S.~S. Pradhan, ``An achievable rate region for the broadcast channel with feedback,'' \emph{IEEE Trans. Inf. Theory}, vol.~59, no.~10, pp. 6175--6191, 2013.

\bibitem{BassiPiantanidaShamai:19p}
G.~Bassi, P.~Piantanida, and S.~Shamai~Shitz, ``The wiretap channel with generalized feedback: Secure communication and key generation,'' \emph{IEEE Trans. Inf. Theory}, vol.~65, no.~4, pp. 2213--2233, 2019.

\bibitem{CaireShamai:99p}
G.~Caire and S.~Shamai, ``On the capacity of some channels with channel state information,'' \emph{IEEE Trans. Inf. Theory}, vol.~45, no.~6, pp. 2007--2019, 1999.

\bibitem{RosenzweigSteinbergShamai:05p}
A.~Rosenzweig, Y.~Steinberg, and S.~Shamai, ``On channels with partial channel state information at the transmitter,'' \emph{IEEE Trans. Inf. Theory}, vol.~51, no.~5, pp. 1817--1830, 2005.

\bibitem{SteinbergWeissman:12p}
Y.~Steinberg and T.~Weissman, ``The degraded broadcast channel with action-dependent states,'' in \emph{2012 IEEE International Symposium on Information Theory Proceedings}, 2012, pp. 596--600.

\bibitem{MerhavWeissman:06p}
N.~Merhav and T.~Weissman, ``Coding for the feedback gel'fand–pinsker channel and the feedforward wyner–ziv source,'' \emph{IEEE Trans. Inf. Theory}, vol.~52, no.~9, pp. 4207--4211, 2006.

\bibitem{CaireShamaiVerdu:05p}
G.~Caire, S.~Shamai, and S.~Verdu, ``An efficient scheme for reliable error correction with limited feedback,'' in \emph{Proc. IEEE Int. Symp. Inf. Theory (ISIT)}, 2005, pp. 1521--1525.

\bibitem{MishraVasalKim:23p}
R.~K. Mishra, D.~Vasal, and H.~Kim, ``Linear coding for {AWGN} channels with noisy output feedback via dynamic programming,'' \emph{IEEE Trans. Inf. Theory}, vol.~69, no.~8, pp. 4889--4906, 2023.

\bibitem{YavasKostinaEffros:24p}
R.~C. Yavas, V.~Kostina, and M.~Effros, ``Variable-length sparse feedback codes for point-to-point, multiple access, and random access channels,'' \emph{IEEE Trans. Inf. Theory}, vol.~70, no.~4, pp. 2367--2394, 2024.

\bibitem{GuptaGuruswamiYunZhang:25p}
M.~Gupta, V.~Guruswami, and R.~Yun~Zhang, ``Binary error-correcting codes with minimal noiseless feedback,'' \emph{IEEE Trans. Inf. Theory}, vol.~71, no.~5, pp. 3424--3446, 2025.

\bibitem{KlimovitchWinter:05a}
G.~V. Klimovitch and A.~Winter, ``Classical capacity of quantum binary adder channels,'' \emph{arXiv preprint quant-ph/0502055}, 2005.

\bibitem{Ding:20p}
D.~Ding, H.~Gharibyan, P.~Hayden, and M.~Walter, ``A quantum multiparty packing lemma and the relay channel,'' \emph{IEEE Transactions on Information Theory}, vol.~66, no.~6, pp. 3500--3519, 2020.

\bibitem{Wilde:11b}
M.~M. Wilde, \emph{Quantum {I}nformation {T}heory}, 2nd~ed.\hskip 1em plus 0.5em minus 0.4em\relax Cambridge University Press, 2017.

\bibitem{Pereg:21p}
U.~Pereg, ``Communication over quantum channels with parameter estimation,'' \emph{IEEE Trans. Inf. Theory}, vol.~68, no.~1, pp. 359--383, 2021.

\bibitem{Fuchs:02b}
C.~A. Fuchs, ``Just two nonorthogonal quantum states,'' in \emph{Quantum Communication, Computing, and Measurement 2}.\hskip 1em plus 0.5em minus 0.4em\relax Springer, 2002, pp. 11--16.

\bibitem{Frohlich:13p}
B.~Fr{\"o}hlich, J.~F. Dynes, M.~Lucamarini, A.~W. Sharpe, Z.~Yuan, and A.~J. Shields, ``A quantum access network,'' \emph{Nature}, vol. 501, no. 7465, pp. 69--72, 2013.

\bibitem{Hirche:18t}
C.~Hirche, ``From asymptotic hypothesis testing to entropy inequalities,'' Ph.D. dissertation, Autonoma Barcelona, Bellaterra, Spain, 2018.

\bibitem{HiaiPetz:91p}
F.~Hiai and D.~Petz, ``The proper formula for relative entropy and its asymptotics in quantum probability,'' \emph{Commun. Math. Phys.}, vol. 143, pp. 99--114, 1991.

\end{thebibliography}

\end{document}